\documentclass{rspublic}
 
\usepackage{epsf}
\usepackage{longtable}
\usepackage{natbib}
\usepackage{epstopdf}
\usepackage{graphicx}

\begin{document} 
 
\title[Coronal Heating]{A Contemporary View of Coronal Heating} 
 
\author[C.E.~Parnell, I.~De Moortel]{Clare E.~Parnell$^1$, Ineke De Moortel$^1$} 
 
\affiliation{$^1$School of Mathematics \& Statistics, University of St Andrews, North Haugh, St Andrews KY16 9SS, UK} 
 
\maketitle 
 
\begin{abstract}{Corona, Chromosphere, Magnetic Fields, Coronal Heating} 
Determining the heating mechanism (or mechanisms) that causes the outer atmosphere of the Sun, and many other stars, to reach temperatures orders of magnitude higher than their surface temperatures has long been a key problem. For decades the problem has been known as the coronal heating problem, but it is now clear that `coronal heating' cannot be treated or explained in isolation and that the heating of the whole solar atmosphere must be studied as a highly coupled system. The magnetic field of the star is known to play a key role, but, despite significant advancements in solar telescopes, computing power and much greater understanding of theoretical mechanisms, the question of which mechanism or mechanisms are the dominant supplier of energy to the chromosphere and corona is still open. Following substantial recent progress, we consider the most likely contenders and discuss the key factors that have made, and still make, determining the actual (coronal) heating mechanism (or mechanisms) so difficult.  
\end{abstract} 
 
\section{Introduction}
The coronal heating problem is one of the longest running solar physics puzzles and is still a highly controversial topic. Considerable progress has been made in modelling possible heating mechanisms, but discriminating amongst these to discover if any one process dominates is extremely difficult to do. 
This is because the corona is not energetically isolated from the other regions of the atmosphere, such as the chromosphere, but instead the whole of the solar atmosphere forms a highly coupled system, with both energy and mass transferred in both directions between the chromosphere and corona through the transition region. The chromosphere has very different plasma properties to the corona and so explaining the heating of the solar atmosphere is now recognised to be considerably more complicated than had been appreciated during much of the twentieth century. 

The coronal heating problem first arose following the work of \citet{Grotrian1939} (erroneously the author on the paper is given as W. Grotian) and \citet{Edlen1942} who discovered that emission lines observed during a total solar eclipse in 1869 were not due to a new element called coronium, but were the result of highly ionised iron. This work demonstrated that the temperature of the corona is in excess of a million degrees Kelvin. In comparison, the temperature near the Sun's surface, the photosphere, is just 6,000 K. At the same time the density falls off by six orders of magnitude from the photosphere to the corona. 

The Sun is not unique in having such a hot outer atmosphere. Indeed, most stars in the cool half of the Hertzsprung-Russell (H-R) diagram have a corona and can be observed in X-rays. Their average temperatures range from 1-45 million degrees Kelvin with the hottest associated with the most rapidly rotating stars. Using observations from the Einstein Observatory, \citet{vaiana1981} was the first to show that most stars, as opposed to just the odd exotic star, have soft X-ray emission and, hence, a corona. \citet{linsky1985} reviewed stars across the H-R diagram with emissions in ultraviolet and X-ray emission lines which are produced by plasma hotter than $10^4$ K. He found that Dwarf stars of spectral type G-M and rapidly rotating subgiants and giants of spectral type F-K in spectroscopic binary systems are most likely to be solar-like in nature, i.e., they are likely to have ``a turbulent magnetic field sufficiently strong to control the dynamics and energetics of their outer atmosphere''. Furthermore, he found that T Tauri stars, other pre-Main-Sequence stars and single giants of spectral type F to early K are also probably solar-like. 

\begin{figure}[ht]
\centering{
\includegraphics[scale=0.112]{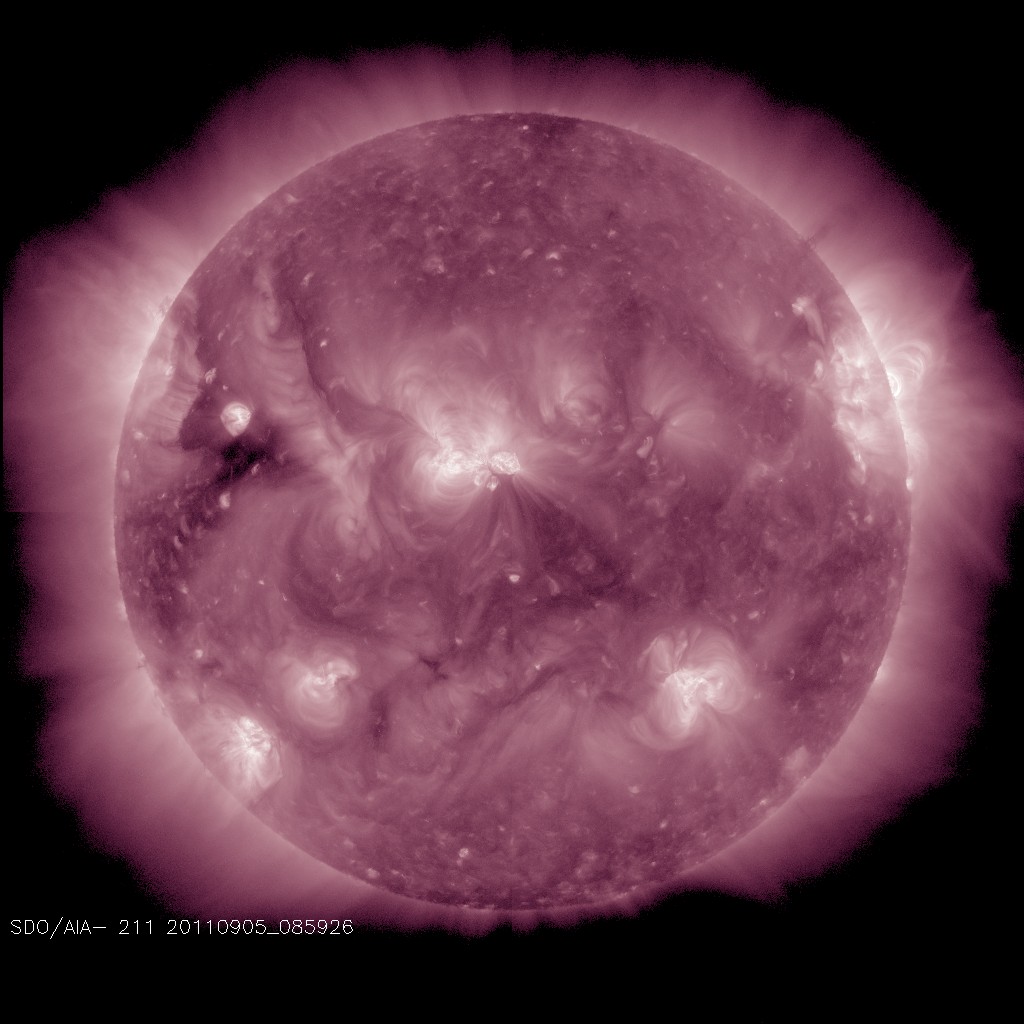}
\includegraphics[scale=0.112]{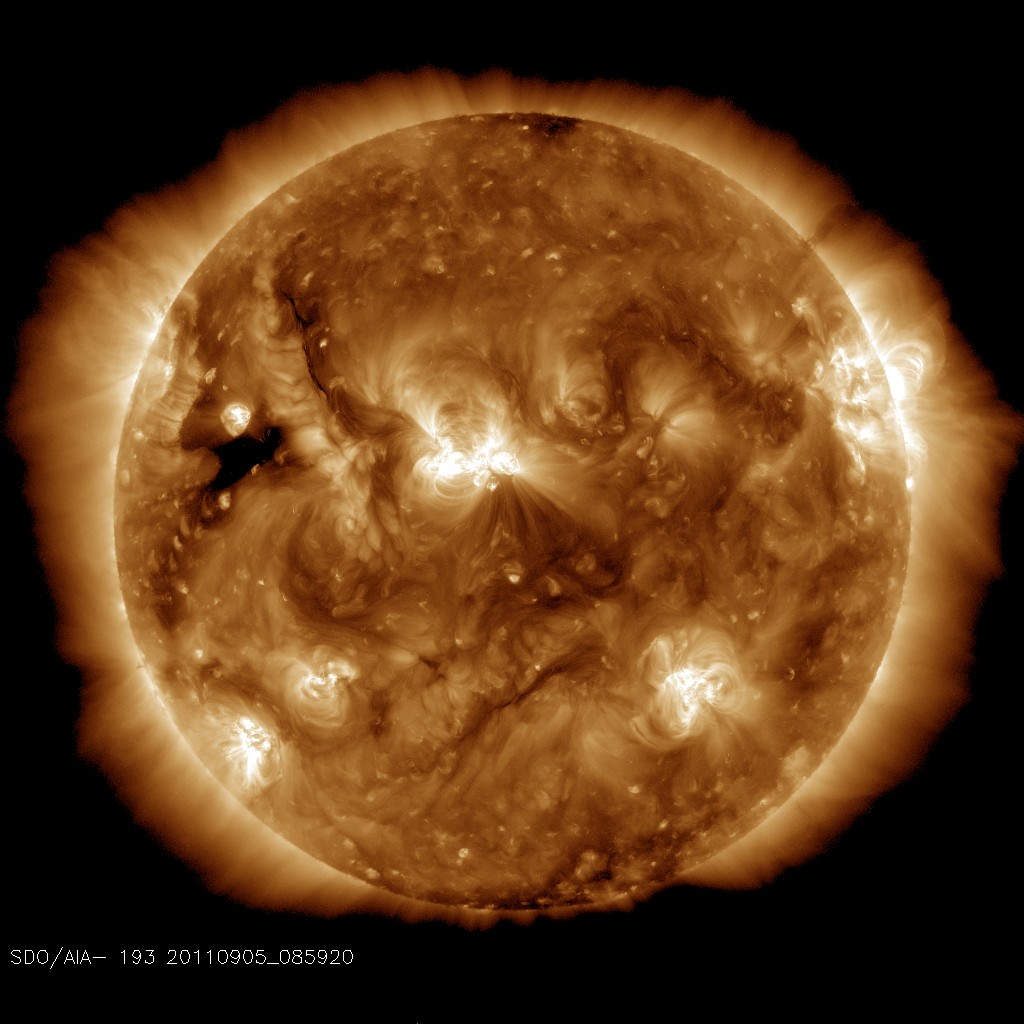}
\includegraphics[scale=0.112]{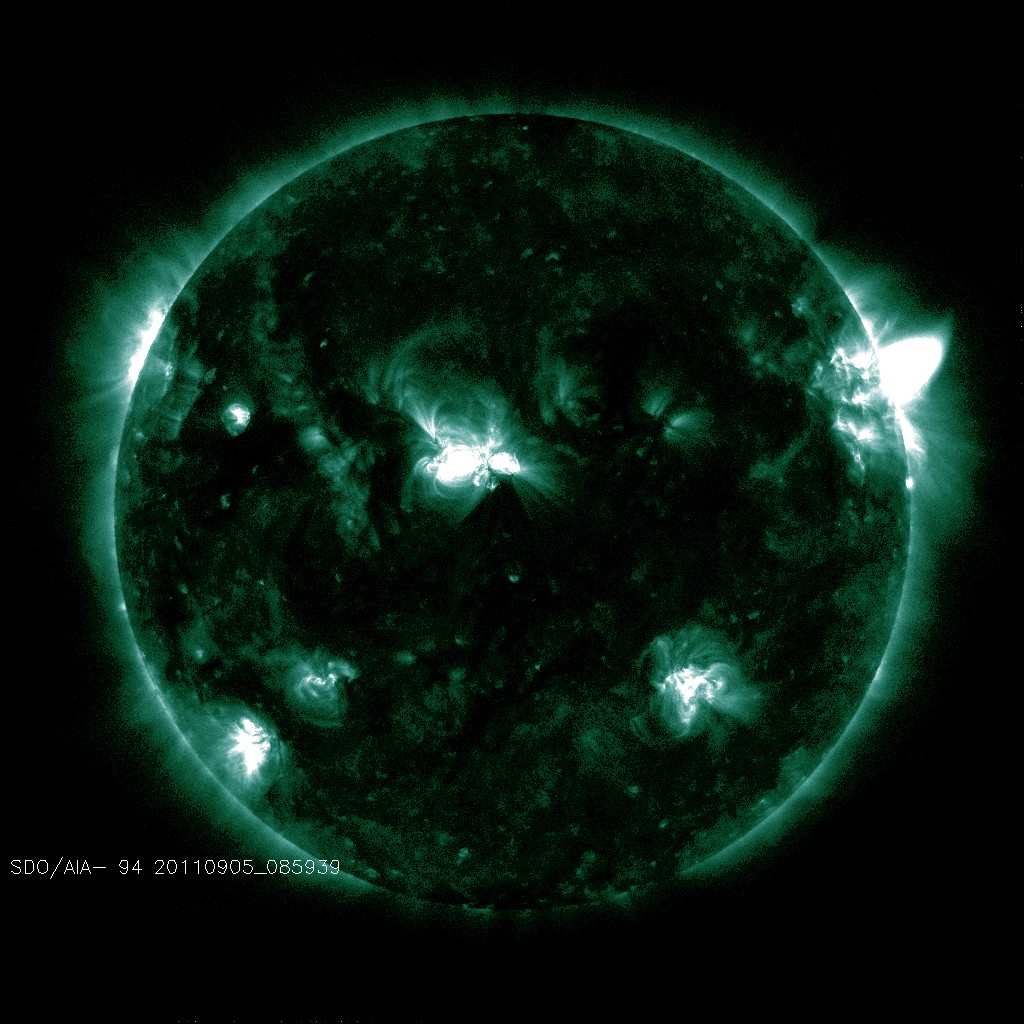}}
\centering{
\includegraphics[scale=0.112]{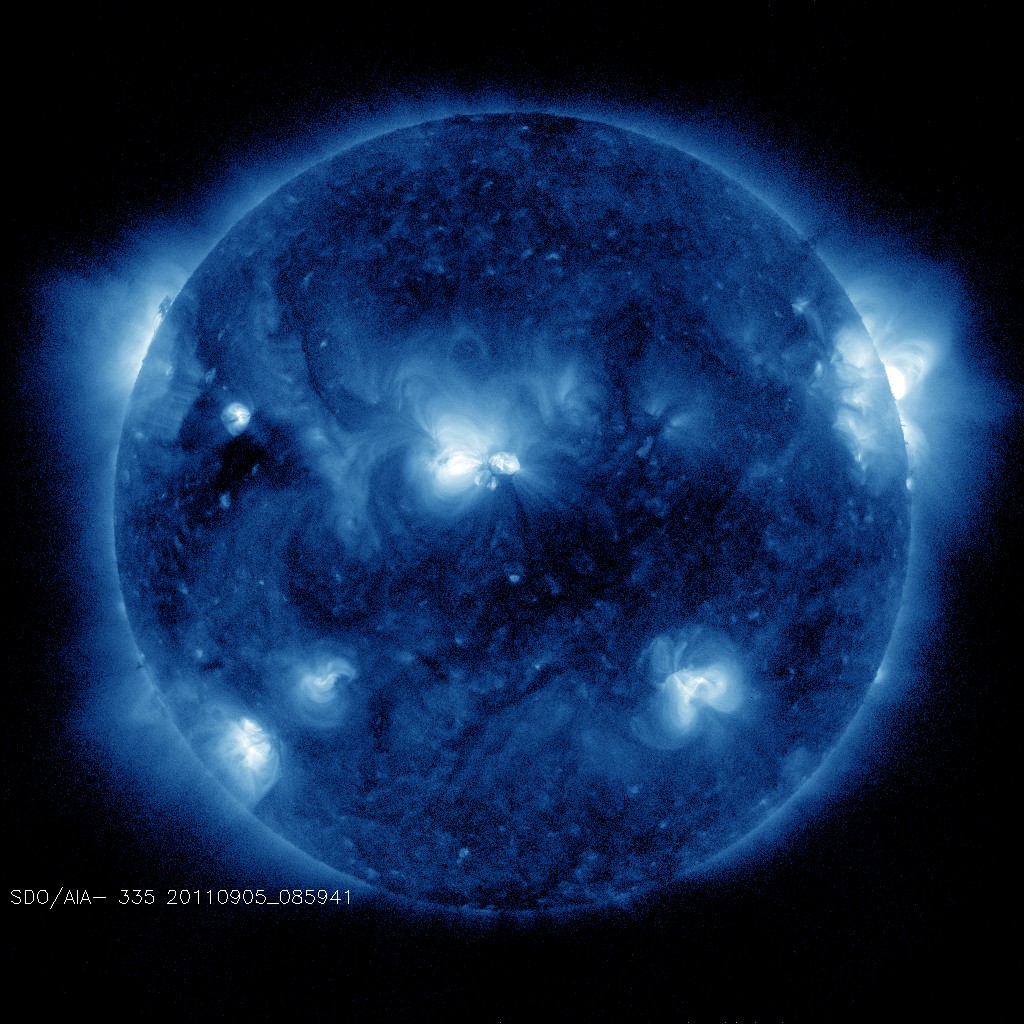}
\includegraphics[scale=0.112]{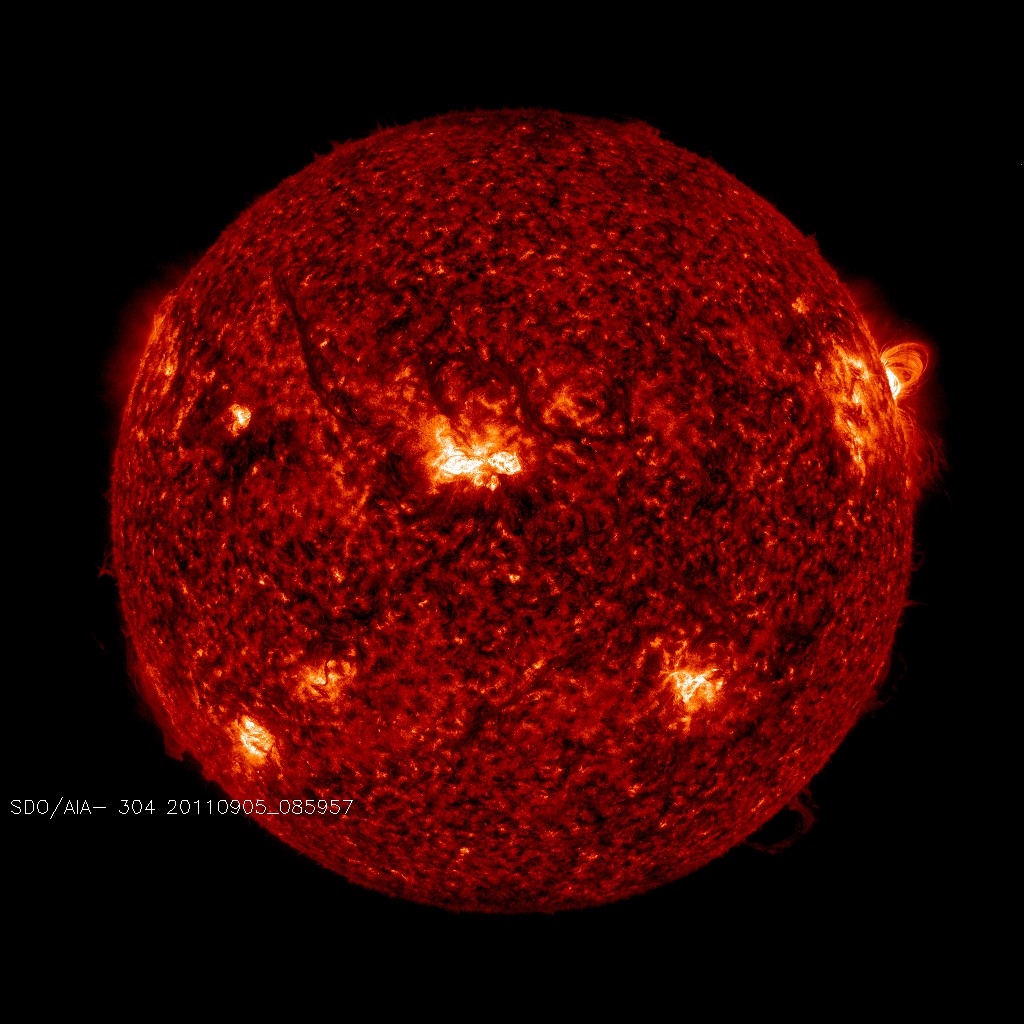}
\includegraphics[scale=0.112]{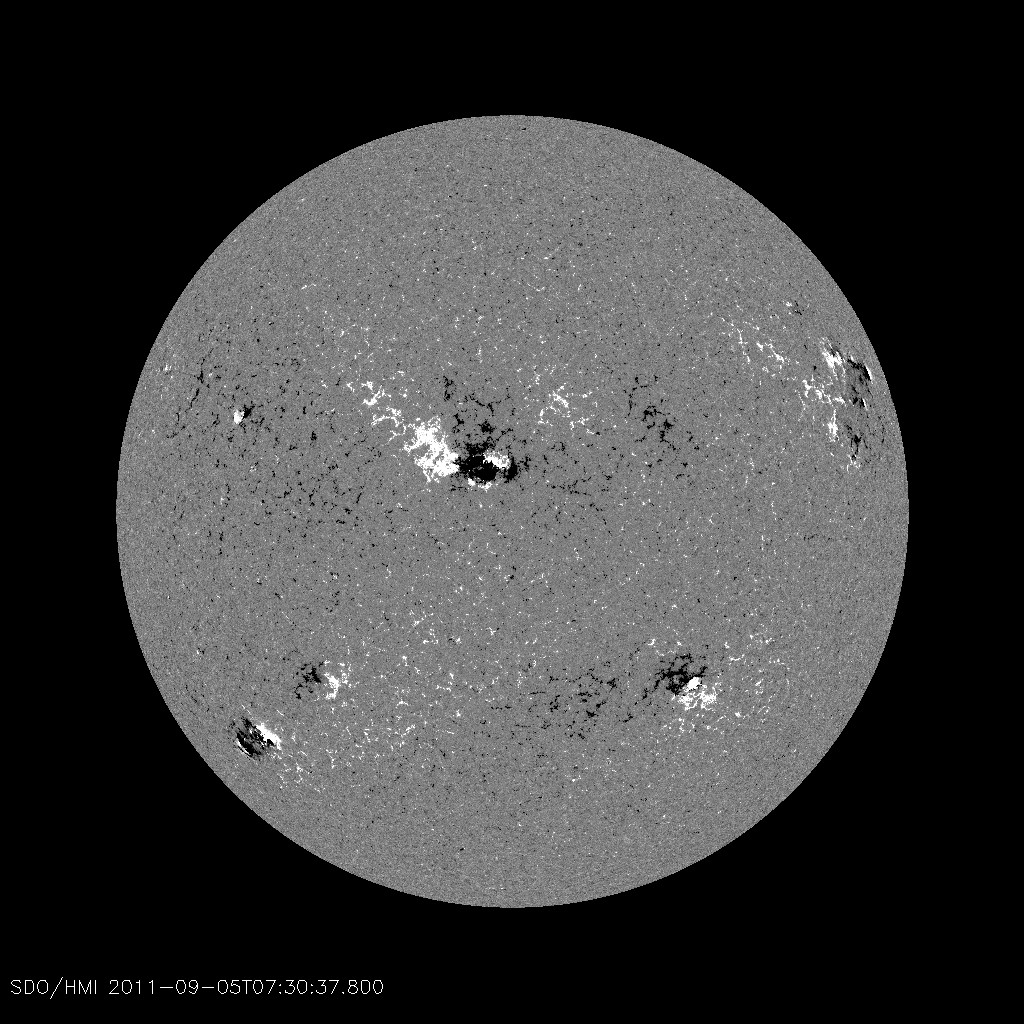}
}
\caption{(a-e) SDO/AIA 211 \AA, 193 \AA, 094 \AA, 335 \AA\ and 304 \AA\ images of the solar corona and chromosphere showing the amazing complexity and vast range of scales of structures. (f) SDO/HMI magnetogram of the magnetic field in the photosphere. All images taken on 27th August 2011. 
\label{fig:sdosun}}
\end{figure}
X-ray, EUV and UV images of the Sun (Figure~\ref{fig:sdosun}(a-e)) show beautiful, finely detailed loops structures, as well as more diffuse emission which is likely to be formed from a mass of unresolved loops. The resolved loops either connect regions of opposite-polarity magnetic field seen in the photosphere, or they are extended open structures that are anchored in a photospheric magnetic feature. As a general rule, the hotter the loops, the stronger the regions of magnetic field they are associated with \citep{Fisher1998}.

 The structure of the magnetic field is highlighted in coronal images because heat is spread efficiently along field lines, but transport across them is greatly inhibited (thermal conduction in the corona is 13 orders of magnitude stronger along field lines than across them \citep{Priest1982}). All of these factors signify that the magnetic field plays a major role in heating the solar corona. Moreover, there is strong evidence from other stars that X-ray coronae are associated with magnetic field since a power-law relation exists between X-ray luminosity and magnetic flux over many orders of magnitude, ranging from solar quiet-regions, through active regions, to dwarf and TTauri stars \citep{Fisher1998,Pevtsov2003}.

The loops observed, for example in Figure~\ref{fig:sdosun}(a-e), are visible since they contain plasma of the appropriate density and temperature to emit at the given wavelength. The surrounding space contains loops that are either at a different temperature  (hotter or cooler) or insufficiently dense to be observed. In equilibrium, and without any heating, a loop will be cool and rarefied, so any heating mechanism must explain both how the material fills the loop and heats it to coronal temperatures. 

In the corona, the combined radiative and conductive losses in active regions are $10^7$ erg cm$^{-2}$ s$^{-1}$ and in the quiet-Sun are $3\times10^5$ erg cm$^{-2}$ s$^{-1}$, whereas in the chromosphere they are $2\times 10^7$ erg cm$^{-2}$ s$^{-1}$ and $4\times 10^6$ erg cm$^{-2}$ s$^{-1}$ in active regions and the quiet-Sun, respectively \citep{Withbroe1977}. The source of this energy is convective churning of the plasma at and below the photosphere. These motions continuously move the footpoints of the ubiquitous magnetic fields that thread through the surface of the Sun with the timescale of these motions, relative to the end-to-end Alfv{\'e}n travel time, leading to one of two fundamental types of heating mechanism. Slow (long timescale) motions result in a quasi-static stressing of the field, whilst fast (short timescale) motions generate waves. The dissipation of magnetic stresses, which typically manifest themselves in the form of current sheets, leads to magnetic reconnection and is known as DC heating, whilst the dissipation of waves is referred to as AC heating. In both cases the actual dissipation will occur at kinetic scales, since the Lundquist number (the ratio of the Ohmic dissipation time to the Alfv{\'e}n crossing time) is very large ($\approx 10^{10-13}$). On the other-hand, the photospheric driving and the complex geometry/topology of the magnetic field are determined globally. Not surprisingly, this massive range of scales which need to be addressed by coronal heating models poses a problem for both theorists and observers. Current parallel computers have neither the memory nor the power to permit well resolved models on all scales and current imagining telescopes have neither the required spatial and temporal resolution nor the sensitivity to see anything at the kinetic scale. Additional difficulties arise in the form of the coupling between the dense interior and tenuous outer atmosphere which require very different physics and, for open magnetic field regions, mechanisms must also account for the mass flux loss due to the solar wind. With all these difficulties it is not surprising that, to date, researchers have been unable to properly explain the solar heating problem. Nonetheless our understanding has advanced significantly.  

Since the identification of the hot solar atmosphere, there have been a huge number of heating mechanisms proposed that involve reconnection or magnetic waves. A full review of the analytical and numerical studies carried out is far beyond the scope of this review, so we refer the interested reader to the reviews by \citet{Aschwanden2004}, \citet{Goedbloed2004}, \cite{Klimchuk2006}, \citet{Aschwanden2007}, \citet{Goossens2011} and \citet{Reale2010}. Instead, we focus on those mechanisms currently most favoured. First, though, in Section 2, we consider the source of the corona's energy, before giving a brief selective discussion on the back grounds to wave heating (Section 3) and magnetic reconnection (Section 4). Nanoflare observations are reviewed in Section 5 and the heating of coronal loops, including numerical modelling and the recent observational results on spicules and Alfv\'enic waves, is the focus of Section 6. Finally, in Section 7, we present our conclusions.

\section{Poynting Flux}
The energy transported into or out of the solar atmosphere is provided by a Poynting flux through the surface of the Sun,
$$S = \frac{1}{\mu_0}\int\int_{S_{\circ}} {\bf E}\times {\bf B}\cdot {\bf dS}\;,$$
where ${\bf E}$ is the electric field, ${\bf B}$ the magnetic field, $\mu_0$ the magnetic permeability and $S_{\circ}$ the solar surface. 
By assuming that motions are ideal and by writing the magnetic field and velocity in terms of components that are parallel or perpendicular to the solar surface (denoted by subscripts $p$ or $n$, respectively) the Poynting flux per unit area through the solar surface can be written as
$$\frac{1}{\mu_0}\left({\bf E}\times{\bf B}\right)\cdot\hat{\bf n} = -\frac{1}{\mu_0}({\bf B}_p\cdot{\bf v}_p)B_n+\frac{1}{\mu_0}({\bf B}_p\cdot{\bf B}_p)v_n\;,$$
where $\hat{\bf n}$ is the normal to the solar surface directed radially outwards.
The first term corresponds to the movement of existing magnetic footpoints by horizontal convective motions whereas the latter represents the emergence or submergence of magnetic flux.

Typically the magnetic field is tilted back away from the direction of motion and so $({\bf B}_p\cdot{\bf v}_p)B_n/\mu_0\approx B_pv_pB_n/\mu_0$ provides an inflow of energy into the solar atmosphere (irrespective of the sign of $B_n$). If, instead the field was titled forward towards the direction of motion, then the Poynting flux would would instead remove energy from the solar atmosphere. Magnetic fields are thought to be held in intense kilo Gauss flux tubes and are moved about with a typical local speed of 1 km s$^{-1}$. Assuming that the magnetic field is inclined at an angle of 20 degrees, and that the filling factor is 1\% in the quiet-Sun (giving a mean field strength of 10 G), then the Poynting flux per unit area is about $2.5\times 10^{7}$ erg cm$^{-2}$ s$^{-1}$. Active regions have much higher filling factors of around 50\% (implying a mean field strength of 500 G) which gives a Poynting flux per unit area of $1.2\times 10^{9}$ erg cm$^{-2}$ s$^{-1}$. In both cases simply considering the energy injected by moving the footpoints seems to give more than enough energy to heat both the chromosphere and corona, even if we assume that half the motions result in a loss rather than a gain in energy. Of course, though, energy may also be required to raise plasma from the chromosphere into the corona, accelerate particles and drive the solar wind in open field regions. With this in mind we also estimate the energy contribution from the emergence of flux.

To obtain an inflow of energy to the solar atmosphere from the second term we require an upflow, and hence the emergence of flux. Newly emerging bipolar regions of flux become highly shredded as they rise through the top layers of the convection zone into the photosphere. They appear as a host of small-scale loops that are aligned so that after emergence they form a bipole with anything up to $\approx 5\times10^{19}-10^{20}$ Mx of flux. Larger bipoles are found, but these are associated with active regions. The Poynting flux associated with the emergence of a single kilo Gauss intense flux tube is $2\times10^{12}$ erg cm$^{-2}$ s$^{-1}$, assuming an upflow of 2 km s$^{-1}$. However, the total emergence rate of new flux is just 450 Mx cm$^{-2}$ day $^{-1}$ \citep{Thornton2011} suggesting a filling factor of $6\times 10^{-7}$ and so, in the quiet Sun, the Poynting flux due to the emergence of new magnetic flux is around $10^{6}$ erg cm$^{-2}$ s$^{-1}$. This is a substantial fraction of the $4.3\times10^{6}$ erg cm$^{-2}$ s$^{-1}$ that is required, but is a factor 25 less than that from the moving of footpoints. In active regions, the Poynting flux from moving flux is similarly dominant. 

The above provides a very simple order of magnitude argument, but its results are qualitatively in agreement with the results from the sophisticated MHD modelling of the upper convection zone to the lower corona using radiative transfer by \cite{AbbettFisher2011}. It is clear that the energy that heats the solar atmosphere comes from the convection zone and this energy manifests itself both as waves and, through the slow stressing, current sheets leading to reconnection. Which of these is most important for chromospheric and coronal heating is still very much an open question.

\section{Acoustic and MHD Waves}
 Almost immediately after the identification in the early 1940's of highly ionised spectral lines, such as FeIX and CaXIV, mechanisms based on the dissipation of acoustic waves were proposed \citep{Biermann1946,Biermann1948,Schwarzschild1948}. However, a heating theory based on waves must overcome several obstacles, namely: (i) waves must be generated in (or below) the solar surface layers, (ii) sufficient energy flux has to be transported as only a fraction of the wave energy will be transmitted into the corona and (iii) the waves have to dissipate (very) efficiently to convert the energy flux into heat in the right place, at the right time. The first of these aspects (i) is probably the easiest to address. Indeed, solar convection generates a mixture of upward propagating waves with an energy flux of several times $10^7$ erg cm$^{-2}$  s$^{-1}$\citep{Narain1996}, which would be more than adequate to heat the solar corona (and accelerate the solar wind). However, (ii) is not so easy to satisfy since most of the (magneto)acoustic waves do not propagate into the corona due to strong reflection and refraction off the rapid density and temperature increase in the Transition Region (in the traditional, plane-parallel view of the solar atmosphere). Additionally, mode coupling is expected to take place in regions where the sound and Alfv\'en speed are approximately equal (see e.g.~the numerical simulations by \cite{Rosenthal2002} and \cite{Bogdan2003}) reducing the flux reaching the corona still further. Alfv\'en waves, on the other hand, are generally seen as very efficient carriers of energy over large distances \citep{Hollweg1978,Hollweg1984,Kudoh1999}. With regard to issue (iii), although \citet{Wentzel1974,Wentzel1976} pointed out that Alfv\'en waves could heat the corona, their weak damping makes them problematic as a coronal heating mechanism. This led to the development of a variety of mechanisms which are likely to enhance the dissipation of Alfv\'en waves such as resonant absorption \citep{Ionson1978,Rae1981,Goossens1992,Ofman1998} and phase mixing \citep{Heyvaerts1983}. Both resonant absorption and phase mixing are based on the property that in an inhomogeneous plasma, individual surfaces can oscillate with their own Alfv\'en frequency. In resonant absorption, on a specific magnetic surface, an incoming wave motion can be in resonance with local oscillations, allowing the transfer of energy from large-scale motions to small lengthscales where dissipation can become effective. Similarly, phase mixing describes how Alfv\'en waves on neighbouring field lines (with different Alfv\'en speeds due to a transverse inhomogeneity in the density and/or the magnetic field strength) gradually become out of phase, again generating increasingly large transverse gradients until dissipative lengthscales are reached. Although mostly treated separately in the literature, examples of the close interplay between resonant absorption and phase mixing can be found in e.g.~\citet{Ruderman1997a,Ruderman1997b}. Apart from resonant absorption and phase mixing, other mechanisms such as (non-linear) mode conversion (to modes which might more readily dissipate) and ion-cyclotron resonance have also been proposed (see for example, reviews by \citet{Aschwanden2004,Goedbloed2004,Klimchuk2006,Erdelyi2007} and references therein).

Although many wave-based heating theories were developed, especially in the 1980's and early 1990's, \citep[See][for more detailed reviews]{Aschwanden2004, Goedbloed2004,Goossens2011} reconnection-based mechanisms were generally favoured to address the coronal heating problem. \citet{Porter1994} point out that observed non-thermal line broadenings imply sufficient wave fluxes to account for the energy requirements of both quiet Sun and active regions, but that this does not necessarily mean that these fluxes are readily converted into heat. Indeed, even if the line broadenings are actually produced by waves, the efficiency of many of the theoretically studied wave-based heating mechanism relies on the presence of waves of the right type and the right frequency, which, {\it at the time,} could not be verified since direct observations of waves in the solar atmosphere were largely absent. This situation has changed with the advent of space-based EUV imagers (SOHO/EIT and TRACE), which now have sufficiently high spatial and temporal resolution to resolve oscillatory motions. A large variety of waves and oscillations have since been observed in the solar atmosphere. However, the {\it observed} energy budget contained within the observed waves was generally several orders of magnitude too small to account for the heating of coronal loops. Additionally, given the rapid damping of many of the observed oscillations, it is not at all clear that the timescales involved even allow for heating to occur during the actual oscillations. Hence, wave-based coronal heating mechanisms initially gained little ground. The discovery of several types of MHD waves in the solar coronal did however led to the rapid development of coronal seismology \citep[see the review by][in this issue]{DeMoortel2012a}. 

More recently observations of flows and Alfv{\'e}n waves have led to a renewed interest in wave heating mechanisms (see Section 6c and 6d), both observationally and theoretically. For example, \cite{VanDoorsselaere2007} look at coronal loop heating profiles and find that, based on observational evidence, a resistive wave heating mechanism is more likely than a viscous one. Similarly, \cite{Taroyan2007} show that power spectra of Doppler shift oscillations might allow a distinction between models based on footpoint heating and uniform heating. A recent study by \cite{Antolin2010} investigates heating based on torsional Alfv\'en waves and suggests that, within the constraints of their 1.5D model, this mechanism is only viable in a rather narrow regime, requiring long and thick loops (see \cite{Jess2009} for a possible observational identification of torsional Alfv\'en waves in the solar chromosphere). We refer the interested reader to \cite{Taroyan2009} for a review of heating diagnostics with MHD waves.

\section{Magnetic Reconnection}
One consequence of new flux ($\gtrsim 10^{19}$ Mx) emerging into the photosphere is the formation of new coronal loops connecting the new flux features to pre-existing ones. The only way these new loops could have formed is through magnetic reconnection (see article by Pontin in this volume or books by \citet{Priest2000} and \citet{Biskamp2000} for more details) allowing a global restructuring of the magnetic field. The removal of flux from the photosphere is also associated with magnetic reconnection since previously unconnected pairs of magnetic features must connect before they can cancel. In both cases, the reconnection driven by emergence or cancellation of flux may be associated with not only the formation and heating of new loops, but also the bulk acceleration of plasma, particle acceleration and the lifting of mass into the chromosphere/corona \citep[e.g.][]{Archontis2005,vonRekowski2008,Martinez-Sykora2009}. Depending on the location and rate of reconnection and also the amount of plasma lifted, such processes may result in loops that can range from hot coronal loops to cool dense chromospheric loops. 

Emerging and/or cancelling flux models have been used to explain X-ray bright points \citep{Heyvaerts1977,Priest1994,Parnell1994,Longcope1998} and X-ray jets \citep{Shibata1992a,Shibata1992b,Yokoyama1995,Moreno-Insertis2008}, although, the basic principles of these models also work on larger scales in active regions \citep[e.g.][]{Masuda1994,Archontis2005,Martinez-Sykora2009,Parnell2010} indicating that many of the observed coronal loops, large and small, may well result from reconnection. 

The processes of emergence and cancellation, however, are not essential for reconnection to occur. The mere movement of magnetic footpoints can also lead to reconnection \citep{Longcope1998,Galsgaard2000,Aulanier2006,DeMoortel2006, Haynes2007,Aulanier2007}. \citet{Close2004} used a magnetic feature tracking algorithm to identify and track magnetic features observed in quiet-Sun MDI high-resolution magnetograms. They then extrapolated the magnetic field into the solar atmosphere and quantified the changes in magnetic connections between features as the photospheric magnetic field evolved. Their work suggested that all the flux in the corona is recycled (all the connections are broken and reconnected) in around 1.4 hrs, a rate at least ten times faster than the recycling time of the photospheric flux as observed by MDI \citep{Hagenaar2001}. Furthermore, since the flux in the photosphere is contained predominantly within small-scale magnetic features, the complexity in the structure of the magnetic field falls off rapidly, with more than 50\% of the flux closing down below 2.5 Mm and just 10\% reaching above 25 Mm \citep{Close2003}. These results indicate that the majority of the magnetic interactions (reconnections) between loops will occur below these heights, quite probably in the chromosphere. 

The models of \citep{Close2003,Close2004} are oversimplified and more sophisticated models are required to quantify these effects more accurately. For instance, in the solar atmosphere reconnection occurs more readily in the cool dense chromospheric region than in the rarefied hot corona since diffusivity is inversely proportional to temperature.       
Furthermore, reconnection will convert magnetic energy into both thermal and kinetic energy (generating waves and shocks), as well as particle acceleration. In particular, there will be direct heating from ohmic dissipation of the plasma in field lines threading the reconnection site, as well as indirect heating of the surrounding plasma due to the viscous damping of waves, outflow jets and shocks created by the reconnection. In some reconnection models, e.g., \cite{Petschek1964} it is believed that the majority of the heating comes from the shocks. Although, recent numerical experiments suggests ohmic heating dominates, in both the chromosphere and corona, for some reconnection scenarios \citep{FuentesFernandez2012a,FuentesFernandez2012b}. Thus it is still an open question as to whether viscous or Ohmic heating mechanisms dominate and what the location of the energy deposition is relative to the reconnection site. However, it is clear that the precise nature will depend on the particular reconnection regime operating and the properties surrounding plasma.

\citet{Gudiksen2005b,Gudiksen2005a} were the first to run 3D MHD models solving an energy equation that included field aligned thermal conduction, as well as radiative losses. Their results qualitatively match coronal observations in many ways, but in such complex numerical experiments it was not possible, at the time, to identify exactly which physical processes were responsible for the heating. For instance, some key questions which need to be addressed are, in their model is reconnection the key heating mechanism and if so is the reconnection occurring principally in the corona or the chromosphere? Also, the cadence of their experiments was such that analysing the role of waves was not possible. Yet more sophisticated modelling \citep[e.g.,][]{Martinez-Sykora2009,Hansteen2010}, including additional physics such as optically thick radiation in the chromosphere, has also been performed. These models find episodic heating events in the chromosphere and it is claimed that these events are due to `stressing of the magnetic field' by photospheric motions, as well as flux emergence, but a detailed analysis on the true nature of these events has yet to be undertaken.

\section{Nanoflares - Observations}
One important heating mechanism, the basics of which were proposed almost forty years ago, is the dissipation of a multitude of small-scale current sheets \citep{Parker1972,Levine1974}. In particular, \citet{Parker1972} hypothesised that slow convective motions of magnetic foot points could lead to a braiding of the magnetic field, creating small current sheets which would be continually dissipated and reformed, providing direct heating via Ohmic dissipation. This concept may be thought of as a braiding between distinct magnetic loops or a self braiding of the magnetic field within a single loop. In either case, magnetic loops need not necessarily wrap around each other, but simple shearing motions suffice to produce current sheets.

 Following the discovery by \citet{Lin1984} from rocket-borne observations of many small hard X-ray spikes with energies between $10^{24}-10^{27}$ ergs, \citet{Parker1988} speculated that a sea of small flares (`nanoflares'), with energies nine orders of magnitude less than large flares (i.e. $10^{24}$ ergs), could heat the corona if they occurred in sufficient numbers. This is known as Parker's braiding or nanoflare theory.

 In reality a wide range of events (sizes and energies) have been observed in the solar atmosphere following power-law distributions of frequency with varying indices.
\citet{Hudson1991} considered the implications of a power-law distribution of energies. 
His important paper pointed out that the index of the power-law distribution of energies is critical in establishing whether small-scale or large-scale flares provide more heat. If the powerlaw index $\alpha>-2$, large flares dominate the heating of the corona, whereas if $\alpha<-2$, small flares dominate.  Note, though, even if the slope is steep enough, nanoflares can only heat the corona provided there are enough of them to produce sufficient energy. 

Using HXRBS/SMM, Yohkoh/SXT, SOHO/EIT and TRACE the distribution of flares, microflares and/or nanoflares were determined by many \citep{Crosby1993,Shimizu1995,Krucker1998,Parnell2000,Aschwanden2000,Benz2002,Aschwanden2002}. Power-law indices for the distribution of event energies were reported ranging from $-2.7 < \alpha < -1.5$ leaving the validity of nanoflares as a heating mechanism completely open. Similarly, power-law indices in the range $-2.5 < \alpha < -1.85$ have been reported for stellar flares observed on a range of late-type cool stars (spectral type F-M) \citep{Audard2000,Gudel2003,Arzner2007}.

Observational determinations of $\alpha$ come from the analysis of images taken by different instruments (e.g., SOHO/EIT, TRACE), with different detection routines (e.g., local peaks in emission or intensity in particular lines in a specific order of some specified size in area and peak), different energies (e.g., thermal energy or radiative/conductive losses) determined and different assumptions made. 
\citet{Parnell2004} examined the effects of making different, but reasonable, assumptions to detect flares, determine their energies and calculate the resulting slope of the power law. The approach involved using just two data sets, one of the quiet Sun and one of an active Sun taken by TRACE.

In the end \citet{Parnell2004} determined 1200 different estimates of the power-law index by considering different sets of the possible parameters as used by others in their analysis.
For both data sets (quiet-Sun and active-region) $\alpha$ lay in the range $-2.5$ to $-1.6$ with a mean of $-1.9$. The most reasonable conclusion that one can draw from these results is that it is not possible to accurately determine the distribution of individual nanoflare/flare energies using the method of direct detection. Instead, alternative methods that attempt to detect ensembles of nanoflares have been sought, \citep[e.g.][]{Sakamoto2008, Terzo2011} providing support, but not conclusive observational proof, that nanoflares heat the corona. Thus, nanoflare heating is currently still regarded as one of the most promising coronal heating mechanisms for both coronal loops and the background quiet-Sun.

\section{Heating of Coronal Loops}
Figure~\ref{fig:sdosun} clearly shows that the solar corona is made up of many loops which range from just a few Mm in length to hundreds of Mm. Even the diffuse emission seen in many of the images in Figure~\ref{fig:sdosun} is highly likely to be made up of unresolved mini loops. The footpoints of coronal loops are not made up of single monolithic magnetic features of one polarity, but are made up of many tiny intense flux tubes, or strands, which are continuously being sheared, rotated and compressed as a result of the underlying granular and supergranular motions. Modelling has shown that any one of these motions can create tiny current sheets within a coronal loop, i.e., as in Parker's braiding/nanoflare \citep{Parker1988}, the kink instability model of \citet{Hoodetal2009} or the tectonics idea of \citet{Priest2002}. The idea is that, if there are enough current sheets distributed throughout the loop cross-section, and these are dissipated and regenerated at an appropriate rate, then enough energy may be provided to maintain the loop at more than a million K for several hours, as observed. However, it is also possible that small-scale heating events may result from a range of different wave dissipation mechanisms such as phase mixing \citep{Heyvaerts1983} and resonant absorption \citep{Ionson1978,Hollweg1988}.
 
\subsection{3D MHD Loop Modelling}
\begin{figure}[ht]
\centering{
\includegraphics[scale=0.27]{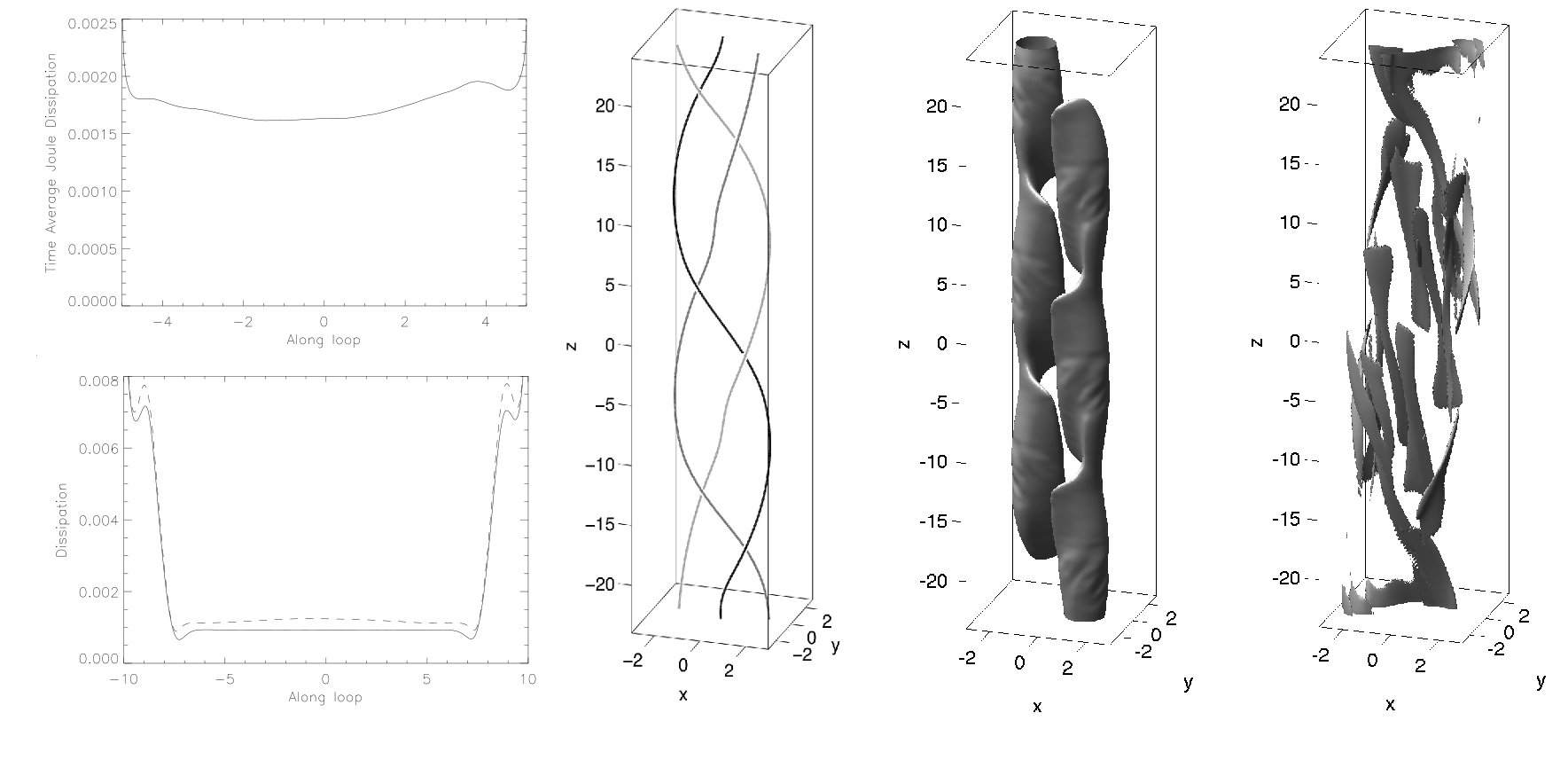}
}
\caption{
The deposition of energy vs position along a curved loop  due to the braiding of a uniform magnetic field in a loop with a (a) uniform atmosphere and (b) a stratified atmosphere (from \citet{Galsgaard2002}). (c) Sample magnetic field lines showing the initial force-free field in the braiding experiment by \citet{Pontin2011}. (d) and (e) show isosurfaces at the start and part way through once reconnection is underway showing how the large current structures have fragmented into a multitude of small-scale sheets providing widespread heating throughout the loop (from \citet{Pontin2011}). 
\label{fig:pontin2011}}
\end{figure}

Models in which simple magnetic fields have been shown to produce multiple current sheets through random driving motions have been conducted by many \citep[e.g.,][]{Mikic1989,Galsgaard1996,Rappazzoetal2007,Rappazzoetal2010}. For instance, \citet{Galsgaard1996} conducted 3D MHD numerical experiments in which they shear an initially uniform vertical magnetic field first in one direction then at right angles. The driving pattern was repeated continuously with random driving periods in each direction forming a quasi-steady state, involving a chaotic distribution of multiple small-scale current sheets throughout the domain where the current sheets continuously dissipate and reform elsewhere. Each small-scale reconnection event may be thought of as a single nanoflare which locally heats a few strands of the coronal loop. However, by averaging over the depth of the domain and over a short time period, the loop appeared to be approximately uniformly heated (Figure~\ref{fig:pontin2011}(a)). If the same braiding experiment is carried out, but this time with a stratified loop, thermal conduction and optically thin radiation, then within the chromospheric regions of the loop considerably more current sheets are both formed and dissipated in comparison to the higher temperature coronal regions (Figure~\ref{fig:pontin2011}(b)) \citep{Galsgaard2002}. This imbalance in the distribution of heating events is due to the significant (a factor of 10) increase in Alfv{\'e}n speed from the photosphere to the corona. The energy density from propagating perturbations is therefore much higher, per unit length, below the transition region than above. Furthermore, the transition region acts as a reflector preventing much of the Poynting flux injected by the boundary motions from reaching the corona.

Instead of a complex driving pattern on a simple magnetic field creating multiple forced reconnection events a multitude of small-scale reconnection events can also be created from a complex magnetic field which undergoes an instability \citep[e.g.,][]{Hoodetal2009,Wilmot-Smith2010,Pontin2011}. \citet{Wilmot-Smith2010} and \citet{Pontin2011} considered the resistive relaxation of a coronal loop consisting of a complex braided magnetic field in which an instability leads to spontaneous reconnection producing a multitude of tiny nanoflares (Figure~\ref{fig:pontin2011}(c-e)). They conducted 3D MHD experiments in which this initially braided magnetic field (containing just 6 crossings), formed by slow footpoint motions, is allowed to relax via resistive MHD. Initially, the magnetic field has large-scale diffuse currents, but, as the field evolves, intense current layers are naturally generated until a resistive instability spontaneously triggers reconnection and the start of a cascade in energy release to small scales. Most importantly they show that in a magnetic loop with what appears to be just 6 reconnection sites (braidings), many hundreds of little energy release events occur, distributed throughout the coronal loop (Figure~\ref{fig:pontin2011}(c)). Moreover, there is much more reconnection than one would first predict. This is because the magnetic field is not clever: it does not follow the route to the quickest and simplest untangling of itself, but rather reconnects wherever the conditions for the onset of reconnection are met \citep{Parnell2010i,Parnell2011}. This results in a small reduction in the magnetic stresses in the field, but not a complete untangling, and a continuous chain of reconnection events must occur before the field can completely untangle itself. Indeed, \citet{Pontin2011} find that the flux in their braided flux tubes can reconnect up to 4.4 times, confirming that the flux is not untangled via the quickest route. Such a process is known as multiple (or recursive) reconnection and was first found by \citet{Parnell2008}. With so many small-scale energy releases throughout the coronal loop, the entire loop can easily be heated and not just a few strands from within it. 

\citet{Parnell2010i} showed that multiple reconnection has a series of important consequences for coronal heating. Not only does it result in a multitude of small-scale reconnection events, and therefore widespread heating throughout the coronal loop, but considerable energy maybe released over a prolonged period. \citet{Parnell2010i} considered an experiment where two opposite-polarity magnetic sources, in a background uniform overlying field, are driven past each other at a constant slow speed of 5\% of the Alfv{\'e}n crossing time. They ran a series of numerical 3D MHD experiments which were exactly the same except for the value of the resistivity. They found that as the resistivity decreased and became more realistic the magnetic reconnection between the opposite-polarity sources was more complex in nature and thus took longer. This resulted in an increase in the horizontal component of the magnetic field at the base and thus an increase in the Poynting flux injected into the system. With more energy in the system, more energy had to be released to get back to the same relaxed state in each case. In their braiding experiment \citet{Pontin2011} note similar findings. Thus, the nature, location and timing of magnetic energy release events are important to understand, not only because of their feedback on the system which may affect the amount of Poynting flux subsequently injected, but also because the plasma/observational consequences will depend greatly on these factors too. 

\subsection{1D Loop Modelling}
Numerical 3D MHD experiments, such as those described above, are very useful in that they can give a good idea as to the distribution, frequency and size of energy release events and can provide an indication of the mechanism(s) heating the corona. However, due to the large global scales they cover they rarely include  all the appropriate physics. In particular, a very simple energy equation is often used neglecting thermal conduction, radiation, radiative transfer and ionization non-equilibrium effects. Thus the plasma response seen in MHD models is not correct and so comparing with observations is extremely difficult. However, the very limited energy transport across fieldlines means that 1D hydro models which can incorporate many of these important physical processes can be used to model the plasma response in a single strand, and a whole coronal loop, by combining the results from many single strand runs \citep[see][for a review]{Klimchuk2006,Reale2010}. 
 
\begin{figure}[t]
\centering{
\scalebox{.4}{\includegraphics{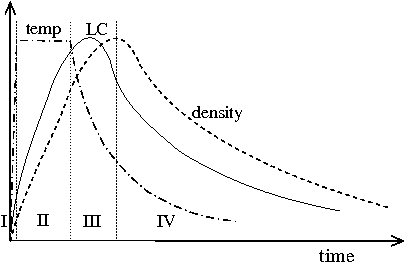}}
\scalebox{.4}{\includegraphics{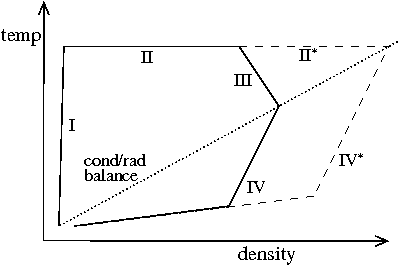}}
}
\caption{(a) Graph depicting the temporal evolution of the temperature (dot-dashed), light curve (solid) and density (dashed) in a loop(strand) due to pulsed heating from a single coronal heating event. (b) A temperature vs density graph of a loop during the same event (Phases I-IV). A long-duration heating pulse may lead to an evolution through phases I, II* and IV*. The dotted line indicates the equilibrium state in which conductive and radiative cooling balance. Graphs based on Figures 14 and 15 from \citet{Reale2010}. 
\label{fig:1Dloopphases}}
\end{figure}
It is highly unlikely that cool dense loops thread the corona waiting to be heated. Instead any coronal heating mechanism must explain both how the plasma is transported into the corona and how it is heated. One crucial result of 1D hydrodynamic models has been to create a clear picture of the behaviour of the plasma within a strand during an impulsive coronal heating event (note that, although often referred to as a `nanoflare', this event may be either AC or DC in origin) from the starting point of an initially cool, low density, equilibrium strand, i.e., one that will not be emitting in EUV or X-rays \citep{Cargill2004, Klimchuk2006,Reale2007}. An impulsive coronal event will cause the strand to evolve through four phases  (Figure~\ref{fig:1Dloopphases}). During the first phase (I), heat from the event is conducted along the field rapidly raising the temperature of the whole strand up to a maximum. The density and emission show very little change during this very short phase creating a strand that is underdense relative to static equilibrium theory. A second longer phase (II), during which the impulsive event continues to heat, follows in which the temperature remains fairly constant, but the chromosphere is strongly heated causing it to expand leading to the evaporation of plasma which rises and fills the strand. During this so-called evaporation phase, the emission rises rapidly and the strand becomes denser, but it still remains underdense. The third phase (III) starts when the impulsive event stops and conduction starts to cool the plasma. The emission first peaks, then starts to decrease, but the density continues to increase. As the strand cools and becomes more dense, radiative cooling effects becomes more important until, at the end of this phase, radiative and conductive cooling finally balance. In phase IV, the loop appears overdense, radiative cooling dominates and the density starts to decrease, slowly at first, until the temperature decreases to a point at which catastrophic cooling occurs and the density plummets until the temperature, density and emission measure return to their pre-event values. 

Clearly, this scenario has a number of key observables that can be tested. Namely, at the start, extremely hot plasma, which, it has been calculated, must be in excess of 4 MK, should be present \citep{Cargill1997}. However, since the loop plasma density is low during this phase, any emission will be very weak and so such hot plasma will be difficult to detect. \citet{Schmeltz2009} have observed hot plasma outside solar flares which is widespread throughout active regions, but weak. A recent paper by \citet{Reale2011} claims to have now found thin high temperature strands of around 10 MK within coronal active regions using SDO/AIA. However, such hot emission is likely to be rare since \citet{Bradshaw2011} argue that nonequilibrium ionization effects will suppress the hot emission lines and in some cases render them invisible. 

Secondly, evaporation upflows should be present. There appears to be a straightforward split with hot chromospheric UV emission lines and coronal EUV lines, in the quiet Sun, with temperatures of formation above 0.5 MK having blue Doppler shifts (e.g., Ne VIII, Mg X, Fe XII), whilst the cooler ($< 0.5$ MK) emission lines (e.g., C II, C IV, Si IV, S VI, N V, O IV, O V, O VI) are red shifted \citep{PeterJudge1999}. In active regions the split is normally higher at around 1 MK. These results may seem consistent with a hot evaporative upflow and a cooler downflow, however, there are other interpretations. In particular, the hot upflows may very well indicate that the actual heating mechanism is not located in the corona, but that plasma in the chromosphere and transition region is heated in situ, resulting in its expansion, creating a flow which fills the loop with hot plasma. There is currently a hot debate over which interpretation is correct, especially since a direct signature of evaporation such as the impact of accelerated particles on the chromosphere is likely to be weak and, hence, very difficult to identify. Most likely though both mechanisms are in operation, but which is most dominant is still an open question.

Finally, since the density behaviour lags that of the temperature during the heating, evaporation and conductive cooling phases (I, II \& III) the strand appears underdense relative to static equilibrium predictions. Observations suggest that hot loops ($T>2$ MK) are underdense \citep{Porter1995},
 and so they are consistent with loops made up of multiple impulsively heated strands where the majority of strands are in Phases I, II and III.
However, so-called warm EUV loops ($T\approx 1-2$ MK) are observed to be overdense relative to equilibrium models and are found to have a narrow temperature range, appearing practically isothermal along their length \citep{Lenz1999,Winebarger2003}. More recently, observations from EIS indicate that the plasma in these loops is not actually isothermal, but rather multithermal with a narrow temperature range ($10^6\pm  3\times 10^5$ K) and a volumetric filling factors of just 10\% \citep{Warren2008a}. Thus it is thought that these warm loops are highly filamentary and cooling from a hotter state \citep{Warren2002,Warren2003}. A possible explanation is that the loop is made up of multiple strands each of which is impulsively heating by a single, short-lived (in comparison to the characteristic cooling time) event. Hence, the plasma in the strand evolves as it is first heated and then cools, such that overall the loop appears to be overdense (in phase IV) for an extended period \citep{Klimchuk2006}. 

For a heating event that is longer than a conductive cooling time the plasma in a strand may follow the phases I, II* and IV* (Figure~\ref{fig:1Dloopphases}b) instead. In fact strands that are heated by either a long slow heating event or a series of repeated short impulsive events could evaporate sufficient density such that the strand reaches a hydrostatic equilibrium and may remain in equilibrium (between phases II* and IV*) if the heating balances the conductive and radiative cooling. When the heating event ceases, radiation is likely to dominate and the temperature and density decrease (Phase IV*) \citep{Klimchuketal2006,Reale2007}. Some loops have a wide range of densities and lengths and, in general, follow a series of scaling laws derived from hydrodynamic equilibrium modelling \citep[such as][]{Rosner1978}. They appear to be consistent with steady heating, i.e., a deposition of heating at a rate that balances the conductive and radiative cooling of the loop \citep[e.g.][]{Porter1995,Kano1995,Warren2010}. Indeed, models of entire active regions or the whole global corona have been made by applying these scaling laws to potential and non-linear force-free magnetic field extrapolations \citep{Schrijver2004,Warren2006,Winebarger2008,Lundquist2008}.

The coronally heated, multistrand scenarios discussed above fit quite a few, but not all observations. In particular, their main problem is that they do not explain how the chromosphere itself is heated, although, if the heating events are due to braiding, then, as the 3D braiding experiments of \cite{Galsgaard2002} seem to suggest, this should not be a problem as an even greater number of heating events should occur in the chromosphere than in the corona. Furthermore, as discussed in the introduction, waves excited by footpoint motions are likely to be abundant in the chromosphere. Nevertheless, a recent series of observations and numerical experiments now suggest that the corona may actually be heated by strong upflows that are driven from the photosphere/chromosphere, as discussed in the next section.

\subsection{Spicules \& Propagating Disturbances} 

\begin{figure}[t]
\centering
\scalebox{.25}{\includegraphics{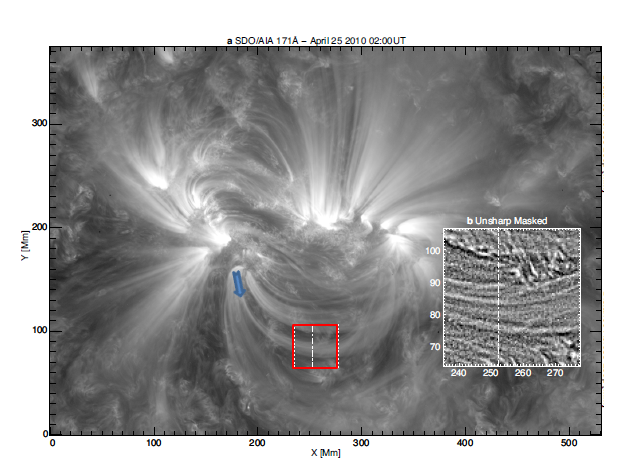}}
\scalebox{.2}{\includegraphics{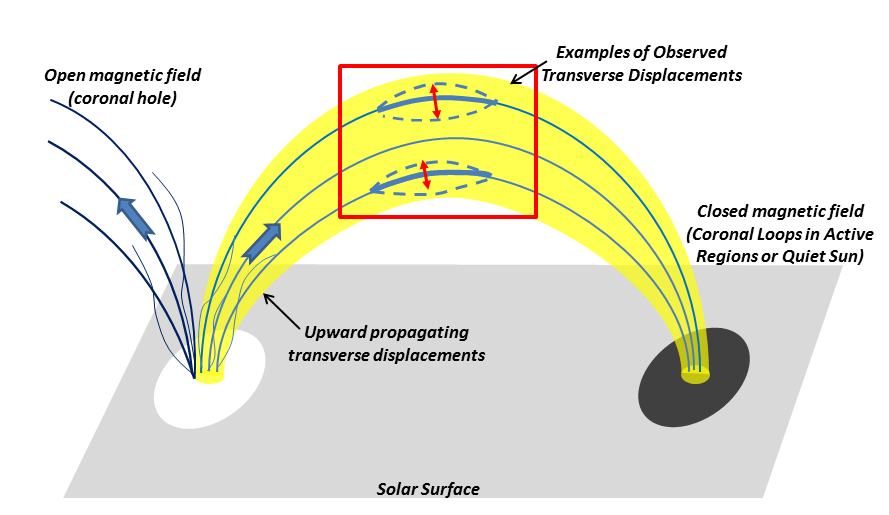}}
\caption{(left) An SDO/AIA 171 \AA\ snapshot of the active region analysed by \cite{McIntosh2011}. Inset is an unsharp-masked image of the region outlined in the red box, showing the transverse loop(thread) displacements. (right) Schematic representation of the footpoint driven Alfv\'enic perturbations in open and closed field geometries (representing coronal holes and quiet-Sun/active-region loops, respectively). The red arrows represent the oscillatory displacements observed by \cite{McIntosh2011} in coronal loops.}
\label{fig:Alfven}
\end{figure}

Following the launch of SOHO and TRACE, several authors such as \citet{Berghmans1999} and \citet{DeMoortel2000}, reported the observation of propagating disturbances in large, quiescent (fan) loops at the edges of active regions, which were interpreted as slow magnetoacoustic waves propagating along coronal loops. However, the energy budget contained within these observed propagating disturbances was generally several orders of magnitude too small to account for the heating of coronal loops.  Even though the coronal energy budget of the propagating slow waves was thought to be very small, \citet{Fontenla1993} and \citet{Jefferies2006} point out that in the lower atmosphere, the energy flux contained in the low-frequency waves ($<$ 5 mHz) is much higher than the flux in their high-frequency counterparts. These high-frequency waves had long been considered as prime candidates to address the heating of the chromosphere, an idea which was rejected by \citet{Fossum2005}. However, their low-frequency counterparts seem to contain much more energy than previously thought.

The potential of the low-frequency propagating disturbances has very much come to the forefront again with recent observational results from Hinode and SDO/AIA as they have been linked with the supply of mass and energy to the solar corona and solar wind. \cite{DePontieu2009} and \cite{DePontieu2011} suggest that a significant fraction of the coronal mass is supplied by chromospheric jets (``Type II spicules''). These authors analyse Hinode and SDO data, in which they find faint, but ubiquitous, quasi-periodic disturbances, interpreted by these authors as flows rather than slow magnetoacoustic waves \citep[see][for more details on this debate]{DePontieu2010,Verwichte2010}, and estimate that the mass contained within these spicules could supply a substantial part of the coronal hot plasma. Recently, \cite{McIntosh2012} reported on observations of slow, counterstreaming downflows ($\sim$10 km/s) in cooler spectroscopic lines, which the authors suggest could be the return-flow of coronal material (i.e.~the end of the mass cycle). Although this mechanism is under considerable debate, it certainly highlights that the corona cannot be treated in isolation. 

\subsection{Transverse, Alfv\'enic Waves}

As well as quasi-periodic propagating disturbances (flows/waves) propagating along coronal loops, recent observations have revealed transverse, propagating oscillations along a variety of structures such as spicules \citep{DePontieu2007,He2009a,He2009b}, X-ray jets \citep{Cirtain2007}, prominence threads \citep{Okamoto2007} and coronal loops \citep{Tomczyk2007,Tomczyk2009,McIntosh2011}. An example of the active region analysed by \cite{McIntosh2011} is shown in Fig.~\ref{fig:Alfven}, who report on small-amplitude transverse displacements in coronal holes, quiet-Sun and active-region loops. An example of the loop(thread) displacements can be seen in the unsharp masked image inset in Fig.~\ref{fig:Alfven} (left). As these footpoint-driven, oscillatory displacements are interpreted as `Alfv\'enic' (see schematic presentation in Fig.~\ref{fig:Alfven}), they are of particular interest to the coronal heating question. Indeed, \cite{DePontieu2007} estimate that the energy flux carried by chromospheric Alfv\'enic waves (observed by Hinode/SOT) is sufficient to accelerate the solar wind and possibly heat the quiet-Sun corona. Although \cite{Tomczyk2007} report that the energy budget estimated from the COMP observations of propagating disturbances in off-limb coronal loops is too small to account for coronal loop heating requirements, \cite{McIntosh2011} not only demonstrate that small-amplitude, transverse, Alfv\'enic perturbations are ubiquitous in the solar corona but that the energy contained in these oscillations appears sufficient to address the requirements of the quiet Sun and coronal holes ($\sim$ $10^5-2\times10^5$ erg cm$^{-2}$ s$^{-1}$). In active regions, the reported energy ($\sim$ $10^5$ erg cm$^{-2}$ s$^{-1}$) is too small to account for the heating of coronal loops (which according to \cite{Withbroe1977} would require about $2\times10^6$ erg cm$^{-2}$ s$^{-1}$). However, in active regions, the superposition of a large number of randomly directed oscillating loop(strands) along the line of sight implies that the estimate reported by McIntosh et al (2011) is likely to be a lower limit. Similarly, a large number of loops, with transverse displacement in random directions is likely to be present off-limb along the line of sight, which could explain why the energy budget estimated by \cite{Tomczyk2007} seemed very small. Using 3D MHD numerical simulations, \cite{DeMoortel2012b} show that the energy budgets estimated from observed, line-of-sight integrated Doppler velocities (such as those observed by \cite{Tomczyk2007}) will be at least an order of magnitude smaller than the energy present in the 3D coronal volume.

Although there is currently still considerable discussion about the interpretation of the propagating disturbances and the transverse displacements, what is clear is that the role of waves in the coronal heating problem has to be reassessed in light of the recent observations by SDO and Hinode. It now appears that Alfv\'enic waves are ubiquitous in the solar atmosphere, potentially containing a substantial amount of energy, and that chromospheric jets form an integral part of the mass supply cycle for the corona and solar wind. However, this does not necessarily imply a solution to the coronal heating problem. Although observations have unambiguously shown that waves are present, theoretical modelling now needs to address how these lead to heating, in the right locations, at the right time.  Indeed, one has to keep in mind that although there appears to be a considerable amount of energy in the observed Alfv\'enic waves \citep[see][]{McIntosh2011}, this does not translate automatically into heating of the plasma. In particular, several authors have recently interpreted the observed damping of, for example, the transverse oscillations observed by COMP \citep{Tomczyk2007,Tomczyk2009} as mode coupling, whereby the amplitude of the footpoint driven, observed, transverse oscillations decreases due to mode coupling to an azimuthal Alfv\'en wave in the tube boundary \citep[see][for a detailed discussion]{DeMoortel2012a}. Potentially, the dissipation of the azimuthal Alfv\'en waves, enhanced by phase mixing in the inhomogeneous tube boundaries could lead to local plasma heating. However, the viability of this remains to be demonstrated.

\section{Discussion \& Conclusions}

For decades after the discovery of the very high temperatures in the solar atmosphere, solving the coronal heating problem was mostly framed as an `AC or DC question', or `waves versus reconnection'. Within the last few years, observational results have shown that the coronal heating problem has to be thought of in a different way. The highly dynamical nature of the solar atmosphere will clearly lead to the occurrence of both types of processes and the true question is to determine the relative contribution of the different mechanisms in different coronal structures. Wave-based heating mechanisms very much seem to have come full circle, from being the first mechanism suggested to provide the necessary energy input into the solar atmosphere, to being mostly discarded and recently back to the forefront again, driven by the spectacular observations by Hinode and SDO. Additionally, the discovery of the abundance of Type II spicules demands that the impact of the equally dynamic chromosphere has to be assessed much more carefully and that it is essential that the `coronal' heating question must be studied as part of the highly coupled solar atmosphere. Indeed, it is well known (but often ignored) that the energy requirements of the chromosphere far exceed those of the corona. 

Heating by multiple small-scale reconnection events has long been considered a viable mechanism and much support has been lent to this mechanism both observationally and theoretically, but definitive observations of the smoking-gun are still missing. Indeed for all coronal heating mechanisms there are still issues that need to be addressed before a definitive conclusion can be reached. 
For instance, the inability to directly `observe' coronal heating has meant that proxies have had to be found. However, it is very difficult using these proxies to distinguish absolutely between wave heating or reconnection mechanisms. For example, the drifting of heating layers in phase mixing/resonant absorption means that such a wave-based mechanism will actually be impulsive in nature and so appear very much like a nanoflare heating event \citep{Ofman1998,Moriyasu2004,Klimchuk2006,Mendoza2006}. Recent studies by Antolin and co-authors (\citet{Antolin2008,Antolinetal2010}) investigate the different observational signatures of coronal heating by either nanoflares or the dissipation of Alfv\'en waves. These authors suggest that coronal rain could be a marker for coronal heating mechanisms, shedding light not only on the spatial distribution of heating events in coronal loops but potentially also on the actual heating mechanism itself.

Additionally, on the theoretical front it is clear that simplified models are not sufficient for comparison with observations. The response of the plasma in the solar atmosphere is very complex and interpreting responses of emission lines is not simple either. Forward modelling of theoretical experiments which have physically realistic effects included is essential, but the analysis of these results must be done with considerable care to determine which are important effects and will take time to complete. In the meantime, the argument over which mechanism(s) heat(s) the corona will continue.   

\begin{acknowledgements} 
IDM acknowledges support from a Royal Society University Research Fellowship and CEP acknowledges the support of STFC through a rolling grant. They would like to thank Dr.\ Brian Welsch for helpful discussions and both anonymous referees for their constructive comments which led to significant improvements in the paper.
\end{acknowledgements} 
 
\bibliographystyle{rspublicnat}

\bibliography{references}

\begin{thebibliography}{146}
\providecommand{\natexlab}[1]{#1}
\expandafter\ifx\csname urlstyle\endcsname\relax
  \providecommand{\doi}[1]{doi:\discretionary{}{}{}#1}\else
  \providecommand{\doi}{doi:\discretionary{}{}{}\begingroup
  \urlstyle{rm}\Url}\fi

\bibitem[{{Abbett} \& {Fisher}(2011)}]{AbbettFisher2011}
{Abbett}, W.~P. \& {Fisher}, G.~H. 2011 {Radiative Cooling in MHD Models of the
  Quiet Sun Convection Zone and Corona}.
\newblock \emph{Solar Phys.}, pp. 206--+.
\newblock (\doi{10.1007/s11207-011-9817-3})

\bibitem[{{Antolin} \& {Shibata}(2010)}]{Antolin2010}
{Antolin}, P. \& {Shibata}, K. 2010 {The Role Of Torsional Alfv{\'e}n Waves in
  Coronal Heating}.
\newblock \emph{Astrophys. J.}, \textbf{712}, 494--510.
\newblock (\doi{10.1088/0004-637X/712/1/494})

\bibitem[{{Antolin} \emph{et~al.}(2008){Antolin}, {Shibata}, {Kudoh}, {Shiota}
  \& {Brooks}}]{Antolin2008}
{Antolin}, P., {Shibata}, K., {Kudoh}, T., {Shiota}, D. \& {Brooks}, D. 2008
  {Predicting Observational Signatures of Coronal Heating by Alfv{\'e}n Waves
  and Nanoflares}.
\newblock \emph{Astrophys. J.}, \textbf{688}, 669--682.
\newblock (\doi{10.1086/591998})

\bibitem[{{Antolin} \emph{et~al.}(2010){Antolin}, {Shibata} \&
  {Vissers}}]{Antolinetal2010}
{Antolin}, P., {Shibata}, K. \& {Vissers}, G. 2010 {Coronal Rain as a Marker
  for Coronal Heating Mechanisms}.
\newblock \emph{Astrophys. J.}, \textbf{716}, 154--166.
\newblock (\doi{10.1088/0004-637X/716/1/154})

\bibitem[{{Archontis} \emph{et~al.}(2005){Archontis}, {Moreno-Insertis},
  {Galsgaard} \& {Hood}}]{Archontis2005}
{Archontis}, V., {Moreno-Insertis}, F., {Galsgaard}, K. \& {Hood}, A.~W. 2005
  {The Three-dimensional Interaction between Emerging Magnetic Flux and a
  Large-Scale Coronal Field: Reconnection, Current Sheets, and Jets}.
\newblock \emph{Astrophys. J.}, \textbf{635}, 1299--1318.
\newblock (\doi{10.1086/497533})

\bibitem[{{Arzner} \emph{et~al.}(2007){Arzner}, {G{\"u}del}, {Briggs},
  {Telleschi} \& {Audard}}]{Arzner2007}
{Arzner}, K., {G{\"u}del}, M., {Briggs}, K., {Telleschi}, A. \& {Audard}, M.
  2007 {Statistics of superimposed flares in the Taurus molecular cloud}.
\newblock \emph{Astron. Astrophys.}, \textbf{468}, 477--484.
\newblock (\doi{10.1051/0004-6361:20066551})

\bibitem[{{Aschwanden}(2004)}]{Aschwanden2004}
{Aschwanden}, M.~J. 2004 \emph{{Physics of the Solar Corona. An Introduction}}.
\newblock Praxis Publishing Ltd.

\bibitem[{{Aschwanden} \& {Parnell}(2002)}]{Aschwanden2002}
{Aschwanden}, M.~J. \& {Parnell}, C.~E. 2002 {Nanoflare Statistics from First
  Principles: Fractal Geometry and Temperature Synthesis}.
\newblock \emph{Astrophys. J.}, \textbf{572}, 1048--1071.
\newblock (\doi{10.1086/340385})

\bibitem[{{Aschwanden} \emph{et~al.}(2000){Aschwanden}, {Tarbell},
  {Nightingale}, {Schrijver}, {Title}, {Kankelborg}, {Martens} \&
  {Warren}}]{Aschwanden2000}
{Aschwanden}, M.~J., {Tarbell}, T.~D., {Nightingale}, R.~W., {Schrijver},
  C.~J., {Title}, A., {Kankelborg}, C.~C., {Martens}, P. \& {Warren}, H.~P.
  2000 {Time Variability of the ``Quiet'' Sun Observed with TRACE. II. Physical
  Parameters, Temperature Evolution, and Energetics of Extreme-Ultraviolet
  Nanoflares}.
\newblock \emph{Astrophys. J.}, \textbf{535}, 1047--1065.
\newblock (\doi{10.1086/308867})

\bibitem[{{Aschwanden} \emph{et~al.}(2007){Aschwanden}, {Winebarger},
  {Tsiklauri} \& {Peter}}]{Aschwanden2007}
{Aschwanden}, M.~J., {Winebarger}, A., {Tsiklauri}, D. \& {Peter}, H. 2007 {The
  Coronal Heating Paradox}.
\newblock \emph{Astrophys. J.}, \textbf{659}, 1673--1681.
\newblock (\doi{10.1086/513070})

\bibitem[{{Audard} \emph{et~al.}(2000){Audard}, {G{\"u}del}, {Drake} \&
  {Kashyap}}]{Audard2000}
{Audard}, M., {G{\"u}del}, M., {Drake}, J.~J. \& {Kashyap}, V.~L. 2000
  {Extreme-Ultraviolet Flare Activity in Late-Type Stars}.
\newblock \emph{Astrophys. J.}, \textbf{541}, 396--409.
\newblock (\doi{10.1086/309426})

\bibitem[{{Aulanier} \emph{et~al.}(2007){Aulanier}, {Golub}, {DeLuca},
  {Cirtain}, {Kano}, {Lundquist}, {Narukage}, {Sakao} \&
  {Weber}}]{Aulanier2007}
{Aulanier}, G., {Golub}, L., {DeLuca}, E.~E., {Cirtain}, J.~W., {Kano}, R.,
  {Lundquist}, L.~L., {Narukage}, N., {Sakao}, T. \& {Weber}, M.~A. 2007
  {Slipping Magnetic Reconnection in Coronal Loops}.
\newblock \emph{Science}, \textbf{318}, 1588--.
\newblock (\doi{10.1126/science.1146143})

\bibitem[{{Aulanier} \emph{et~al.}(2006){Aulanier}, {Pariat}, {D{\'e}moulin} \&
  {Devore}}]{Aulanier2006}
{Aulanier}, G., {Pariat}, E., {D{\'e}moulin}, P. \& {Devore}, C.~R. 2006
  {Slip-Running Reconnection in Quasi-Separatrix Layers}.
\newblock \emph{Solar Phys.}, \textbf{238}, 347--376.
\newblock (\doi{10.1007/s11207-006-0230-2})

\bibitem[{{Benz} \& {Krucker}(2002)}]{Benz2002}
{Benz}, A.~O. \& {Krucker}, S. 2002 {Energy Distribution of Microevents in the
  Quiet Solar Corona}.
\newblock \emph{Astrophys. J.}, \textbf{568}, 413--421.
\newblock (\doi{10.1086/338807})

\bibitem[{{Berghmans} \& {Clette}(1999)}]{Berghmans1999}
{Berghmans}, D. \& {Clette}, F. 1999 {Active region EUV transient brightenings
  - First Results by EIT of SOHO JOP80}.
\newblock \emph{Solar Phys.}, \textbf{186}, 207--229.

\bibitem[{{Biermann}(1946)}]{Biermann1946}
{Biermann}, L. 1946 {Zur Deutung der chromosph{\"a}rischen Turbulenz und des
  Exzesses der UV-Strahlung der Sonne}.
\newblock \emph{Die Naturwissenschaften}, \textbf{33}, 118--119.
\newblock (\doi{10.1007/BF00738265})

\bibitem[{{Biermann}(1948)}]{Biermann1948}
{Biermann}, L. 1948 {{\"U}ber die Ursache der chromosph{\"a}rischen Turbulenz
  und des UV-Exzesses der Sonnenstrahlung}.
\newblock \emph{Zeitschrift f{\"u}r Astrophysik}, \textbf{25}, 161--+.

\bibitem[{{Biskamp}(2000)}]{Biskamp2000}
{Biskamp}, D. 2000 \emph{{Magnetic Reconnection in Plasmas}}.
\newblock Cambridge University Press, Cambridge, UK.

\bibitem[{{Bogdan} \emph{et~al.}(2003){Bogdan}, {Carlsson}, {Hansteen},
  {McMurry}, {Rosenthal}, {Johnson}, {Petty-Powell}, {Zita}, {Stein}
  \emph{et~al.}}]{Bogdan2003}
{Bogdan}, T.~J., {Carlsson}, M., {Hansteen}, V.~H., {McMurry}, A., {Rosenthal},
  C.~S., {Johnson}, M., {Petty-Powell}, S., {Zita}, E.~J., {Stein}, R.~F.
  \emph{et~al.} 2003 {Waves in the Magnetized Solar Atmosphere. II. Waves from
  Localized Sources in Magnetic Flux Concentrations}.
\newblock \emph{Astrophys. J.}, \textbf{599}, 626--660.
\newblock (\doi{10.1086/378512})

\bibitem[{{Bradshaw} \& {Klimchuk}(2011)}]{Bradshaw2011}
{Bradshaw}, S.~J. \& {Klimchuk}, J.~A. 2011 {What Dominates the Coronal
  Emission Spectrum During the Cycle of Impulsive Heating and Cooling?}
\newblock \emph{Astrophys. J. Sup.}, \textbf{194}, 26--+.
\newblock (\doi{10.1088/0067-0049/194/2/26})

\bibitem[{{Cargill} \& {Klimchuk}(1997)}]{Cargill1997}
{Cargill}, P.~J. \& {Klimchuk}, J.~A. 1997 {A Nanoflare Explanation for the
  Heating of Coronal Loops Observed by YOHKOH}.
\newblock \emph{Astrophys. J.}, \textbf{478}, 799.
\newblock (\doi{10.1086/303816})

\bibitem[{{Cargill} \& {Klimchuk}(2004)}]{Cargill2004}
{Cargill}, P.~J. \& {Klimchuk}, J.~A. 2004 {Nanoflare Heating of the Corona
  Revisited}.
\newblock \emph{Astrophys. J.}, \textbf{605}, 911--920.
\newblock (\doi{10.1086/382526})

\bibitem[{{Cirtain} \emph{et~al.}(2007){Cirtain}, {Golub}, {Lundquist}, {van
  Ballegooijen}, {Savcheva}, {Shimojo}, {DeLuca}, {Tsuneta}, {Sakao}
  \emph{et~al.}}]{Cirtain2007}
{Cirtain}, J.~W., {Golub}, L., {Lundquist}, L., {van Ballegooijen}, A.,
  {Savcheva}, A., {Shimojo}, M., {DeLuca}, E., {Tsuneta}, S., {Sakao}, T.
  \emph{et~al.} 2007 {Evidence for Alfv{\'e}n Waves in Solar X-ray Jets}.
\newblock \emph{Science}, \textbf{318}, 1580--.
\newblock (\doi{10.1126/science.1147050})

\bibitem[{{Close} \emph{et~al.}(2004){Close}, {Parnell}, {Longcope} \&
  {Priest}}]{Close2004}
{Close}, R.~M., {Parnell}, C.~E., {Longcope}, D.~W. \& {Priest}, E.~R. 2004
  {Recycling of the Solar Corona's Magnetic Field}.
\newblock \emph{Astrophys. J. Lett.}, \textbf{612}, L81--L84.
\newblock (\doi{10.1086/424659})

\bibitem[{{Close} \emph{et~al.}(2003){Close}, {Parnell}, {Mackay} \&
  {Priest}}]{Close2003}
{Close}, R.~M., {Parnell}, C.~E., {Mackay}, D.~H. \& {Priest}, E.~R. 2003
  {Statistical Flux Tube Properties of 3D Magnetic Carpet Fields}.
\newblock \emph{Solar Phys.}, \textbf{212}, 251--275.

\bibitem[{{Crosby} \emph{et~al.}(1993){Crosby}, {Aschwanden} \&
  {Dennis}}]{Crosby1993}
{Crosby}, N.~B., {Aschwanden}, M.~J. \& {Dennis}, B.~R. 1993 {Frequency
  distributions and correlations of solar X-ray flare parameters}.
\newblock \emph{Solar Phys.}, \textbf{143}, 275--299.
\newblock (\doi{10.1007/BF00646488})

\bibitem[{{De Moortel} \& {Galsgaard}(2006)}]{DeMoortel2006}
{De Moortel}, I. \& {Galsgaard}, K. 2006 {Numerical modelling of 3D
  reconnection. II. Comparison between rotational and spinning footpoint
  motions}.
\newblock \emph{Astron. Astrophys.}, \textbf{459}, 627--639.
\newblock (\doi{10.1051/0004-6361:20065716})

\bibitem[{{De Moortel} \emph{et~al.}(2000){De Moortel}, {Ireland} \&
  {Walsh}}]{DeMoortel2000}
{De Moortel}, I., {Ireland}, J. \& {Walsh}, R.~W. 2000 {Observation of
  oscillations in coronal loops}.
\newblock \emph{Astron. Astrophys.}, \textbf{355}, L23--L26.

\bibitem[{{De Moortel} \& {Nakariakov}(2012)}]{DeMoortel2012a}
{De Moortel}, I. \& {Nakariakov}, V.~M. 2012 {MHD Waves and Coronal Seismology:
  an overview of recent results}.
\newblock \emph{Roy.~Soc.~Phil.~Trans.~A}, p. in this journal.

\bibitem[{{De Moortel} \& {Pascoe}(2012)}]{DeMoortel2012b}
{De Moortel}, I. \& {Pascoe}, D. 2012 {The Effects of Line-of-Sight integration
  on Multistrand Coronal Loop Oscillations}.
\newblock \emph{Astron. Astrophys.}, p. in press.

\bibitem[{{De Pontieu} \& {McIntosh}(2010)}]{DePontieu2010}
{De Pontieu}, B. \& {McIntosh}, S.~W. 2010 {Quasi-periodic Propagating Signals
  in the Solar Corona: The Signature of Magnetoacoustic Waves or High-velocity
  Upflows?}
\newblock \emph{Astrophys. J.}, \textbf{722}, 1013--1029.
\newblock (\doi{10.1088/0004-637X/722/2/1013})

\bibitem[{{De Pontieu} \emph{et~al.}(2011){De Pontieu}, {McIntosh}, {Carlsson},
  {Hansteen}, {Tarbell}, {Boerner}, {Martinez-Sykora}, {Schrijver} \&
  {Title}}]{DePontieu2011}
{De Pontieu}, B., {McIntosh}, S.~W., {Carlsson}, M., {Hansteen}, V.~H.,
  {Tarbell}, T.~D., {Boerner}, P., {Martinez-Sykora}, J., {Schrijver}, C.~J. \&
  {Title}, A.~M. 2011 {The Origins of Hot Plasma in the Solar Corona}.
\newblock \emph{Science}, \textbf{331}, 55--.
\newblock (\doi{10.1126/science.1197738})

\bibitem[{{De Pontieu} \emph{et~al.}(2007){De Pontieu}, {McIntosh}, {Carlsson},
  {Hansteen}, {Tarbell}, {Schrijver}, {Title}, {Shine}, {Tsuneta}
  \emph{et~al.}}]{DePontieu2007}
{De Pontieu}, B., {McIntosh}, S.~W., {Carlsson}, M., {Hansteen}, V.~H.,
  {Tarbell}, T.~D., {Schrijver}, C.~J., {Title}, A.~M., {Shine}, R.~A.,
  {Tsuneta}, S. \emph{et~al.} 2007 {Chromospheric Alfv{\'e}nic Waves Strong
  Enough to Power the Solar Wind}.
\newblock \emph{Science}, \textbf{318}, 1574--.
\newblock (\doi{10.1126/science.1151747})

\bibitem[{{De Pontieu} \emph{et~al.}(2009){De Pontieu}, {McIntosh}, {Hansteen}
  \& {Schrijver}}]{DePontieu2009}
{De Pontieu}, B., {McIntosh}, S.~W., {Hansteen}, V.~H. \& {Schrijver}, C.~J.
  2009 {Observing the Roots of Solar Coronal Heating in the Chromosphere}.
\newblock \emph{Astrophys. J. Lett.}, \textbf{701}, L1--L6.
\newblock (\doi{10.1088/0004-637X/701/1/L1})

\bibitem[{{Edl{\'e}n}(1942)}]{Edlen1942}
{Edl{\'e}n}, B. 1942 {Die Deutung der Emissionslinien im Spektrum der
  Sonnenkorona. Mit 6 Abbildungen.}
\newblock \emph{Zeitschrift f{\"u}r Astrophysik}, \textbf{22}, 30--+.

\bibitem[{{Erd{\'e}lyi} \& {Ballai}(2007)}]{Erdelyi2007}
{Erd{\'e}lyi}, R. \& {Ballai}, I. 2007 {Heating of the solar and stellar
  coronae: a review}.
\newblock \emph{Astronomische Nachrichten}, \textbf{328}, 726--733.
\newblock (\doi{10.1002/asna.200710803})

\bibitem[{{Fisher} \emph{et~al.}(1998){Fisher}, {Longcope}, {Metcalf} \&
  {Pevtsov}}]{Fisher1998}
{Fisher}, G.~H., {Longcope}, D.~W., {Metcalf}, T.~R. \& {Pevtsov}, A.~A. 1998
  {Coronal Heating in Active Regions as a Function of Global Magnetic
  Variables}.
\newblock \emph{Astrophys. J.}, \textbf{508}, 885--898.
\newblock (\doi{10.1086/306435})

\bibitem[{{Fontenla} \emph{et~al.}(1993){Fontenla}, {Rabin}, {Hathaway} \&
  {Moore}}]{Fontenla1993}
{Fontenla}, J.~M., {Rabin}, D., {Hathaway}, D.~H. \& {Moore}, R.~L. 1993
  {Measurement of p-mode energy propagation in the quiet solar photosphere}.
\newblock \emph{Astrophys. J.}, \textbf{405}, 787--797.
\newblock (\doi{10.1086/172408})

\bibitem[{{Fossum} \& {Carlsson}(2005)}]{Fossum2005}
{Fossum}, A. \& {Carlsson}, M. 2005 {High-frequency acoustic waves are not
  sufficient to heat the solar chromosphere}.
\newblock \emph{Nature}, \textbf{435}, 919--921.
\newblock (\doi{10.1038/nature03695})

\bibitem[{{Fuentes Fernandez} \& {Parnell}(2012)}]{FuentesFernandez2012b}
{Fuentes Fernandez}, J. \& {Parnell}, C.~E. 2012 {Consequences of spontaneous
  reconnection at a 2D low-beta current layer}.
\newblock \emph{Phys. Plasmas}, p. in press.

\bibitem[{{Fuentes Fernandez} \emph{et~al.}(2012){Fuentes Fernandez},
  {Parnell}, {Hood}, {Priest} \& {Longcope}}]{FuentesFernandez2012a}
{Fuentes Fernandez}, J., {Parnell}, C.~E., {Hood}, A.~W., {Priest}, E.~R. \&
  {Longcope}, D. 2012 {Consequences of spontaneous reconnection at a 2D
  non-force-free current layer}.
\newblock \emph{Phys. Plasmas}, p. submitted.

\bibitem[{{Galsgaard}(2002)}]{Galsgaard2002}
{Galsgaard}, K. 2002 {Flux braiding in a stratified atmosphere}.
\newblock In \emph{Solmag 2002. proceedings of the magnetic coupling of the
  solar atmosphere euroconference} (ed. {H.~Sawaya-Lacoste}), vol. 505 of
  \emph{ESA Special Publication}, pp. 269--272.

\bibitem[{{Galsgaard} \& {Nordlund}(1996)}]{Galsgaard1996}
{Galsgaard}, K. \& {Nordlund}, {\AA}. 1996 {Heating and activity of the solar
  corona 1. Boundary shearing of an initially homogeneous magnetic field}.
\newblock \emph{J. Geophys. Res.}, \textbf{1011}, 13\,445--13\,460.
\newblock (\doi{10.1029/96JA00428})

\bibitem[{{Galsgaard} \emph{et~al.}(2000){Galsgaard}, {Parnell} \&
  {Blaizot}}]{Galsgaard2000}
{Galsgaard}, K., {Parnell}, C.~E. \& {Blaizot}, J. 2000 {Elementary heating
  events - Magnetic interactions between two flux sources}.
\newblock \emph{Astron. Astrophys.}, \textbf{362}, 395--405.

\bibitem[{{Goedbloed} \& {Poedts}(2004)}]{Goedbloed2004}
{Goedbloed}, J.~P.~H. \& {Poedts}, S. 2004 \emph{{Principles of
  Magnetohydrodynamics}}.
\newblock Cambridge University Press, UK.

\bibitem[{{Goossens} \emph{et~al.}(2011){Goossens}, {Erd{\'e}lyi} \&
  {Ruderman}}]{Goossens2011}
{Goossens}, M., {Erd{\'e}lyi}, R. \& {Ruderman}, M.~S. 2011 {Resonant MHD Waves
  in the Solar Atmosphere}.
\newblock \emph{Space Sci. Rev.}, \textbf{158}, 289--338.
\newblock (\doi{10.1007/s11214-010-9702-7})

\bibitem[{{Goossens} \emph{et~al.}(1992){Goossens}, {Hollweg} \&
  {Sakurai}}]{Goossens1992}
{Goossens}, M., {Hollweg}, J.~V. \& {Sakurai}, T. 1992 {Resonant behaviour of
  MHD waves on magnetic flux tubes. III - Effect of equilibrium flow}.
\newblock \emph{Solar Phys.}, \textbf{138}, 233--255.
\newblock (\doi{10.1007/BF00151914})

\bibitem[{{Grotrian}(1939)}]{Grotrian1939}
{Grotrian}, W. 1939 {Zur Frage der Deutung der Linien im Spektrum der
  Sonnenkorona}.
\newblock \emph{Naturwissenschaften}, \textbf{27}, 214--214.
\newblock (\doi{10.1007/BF01488890})

\bibitem[{{G{\"u}del} \emph{et~al.}(2003){G{\"u}del}, {Audard}, {Kashyap},
  {Drake} \& {Guinan}}]{Gudel2003}
{G{\"u}del}, M., {Audard}, M., {Kashyap}, V.~L., {Drake}, J.~J. \& {Guinan},
  E.~F. 2003 {Are Coronae of Magnetically Active Stars Heated by Flares? II.
  Extreme Ultraviolet and X-Ray Flare Statistics and the Differential Emission
  Measure Distribution}.
\newblock \emph{Astrophys. J.}, \textbf{582}, 423--442.
\newblock (\doi{10.1086/344614})

\bibitem[{{Gudiksen} \& {Nordlund}(2005{\natexlab{\emph{a}}})}]{Gudiksen2005b}
{Gudiksen}, B.~V. \& {Nordlund}, {\AA}. 2005{\natexlab{\emph{a}}} {An AB Initio
  Approach to Solar Coronal Loops}.
\newblock \emph{Astrophys. J.}, \textbf{618}, 1031--1038.
\newblock (\doi{10.1086/426064})

\bibitem[{{Gudiksen} \& {Nordlund}(2005{\natexlab{\emph{b}}})}]{Gudiksen2005a}
{Gudiksen}, B.~V. \& {Nordlund}, {\AA}. 2005{\natexlab{\emph{b}}} {An Ab Initio
  Approach to the Solar Coronal Heating Problem}.
\newblock \emph{Astrophys. J.}, \textbf{618}, 1020--1030.
\newblock (\doi{10.1086/426063})

\bibitem[{{Hagenaar}(2001)}]{Hagenaar2001}
{Hagenaar}, H.~J. 2001 {Ephemeral Regions on a Sequence of Full-Disk Michelson
  Doppler Imager Magnetograms}.
\newblock \emph{Astrophys. J.}, \textbf{555}, 448--461.
\newblock (\doi{10.1086/321448})

\bibitem[{{Hansteen} \emph{et~al.}(2010){Hansteen}, {Hara}, {De Pontieu} \&
  {Carlsson}}]{Hansteen2010}
{Hansteen}, V.~H., {Hara}, H., {De Pontieu}, B. \& {Carlsson}, M. 2010 {On
  Redshifts and Blueshifts in the Transition Region and Corona}.
\newblock \emph{Astrophys. J.}, \textbf{718}, 1070--1078.
\newblock (\doi{10.1088/0004-637X/718/2/1070})

\bibitem[{{Haynes} \emph{et~al.}(2007){Haynes}, {Parnell}, {Galsgaard} \&
  {Priest}}]{Haynes2007}
{Haynes}, A.~L., {Parnell}, C.~E., {Galsgaard}, K. \& {Priest}, E.~R. 2007
  {Magnetohydrodynamic evolution of magnetic skeletons}.
\newblock \emph{Roy. Soc. Lon. Proc. Series A}, \textbf{463}, 1097--1115.
\newblock (\doi{10.1098/rspa.2007.1815})

\bibitem[{{He} \emph{et~al.}(2009{\natexlab{\emph{a}}}){He}, {Marsch}, {Tu} \&
  {Tian}}]{He2009a}
{He}, J., {Marsch}, E., {Tu}, C. \& {Tian}, H. 2009{\natexlab{\emph{a}}}
  {Excitation of Kink Waves Due to Small-Scale Magnetic Reconnection in the
  Chromosphere?}
\newblock \emph{Astrophys. J. Lett.}, \textbf{705}, L217--L222.
\newblock (\doi{10.1088/0004-637X/705/2/L217})

\bibitem[{{He} \emph{et~al.}(2009{\natexlab{\emph{b}}}){He}, {Tu}, {Marsch},
  {Guo}, {Yao} \& {Tian}}]{He2009b}
{He}, J.-S., {Tu}, C.-Y., {Marsch}, E., {Guo}, L.-J., {Yao}, S. \& {Tian}, H.
  2009{\natexlab{\emph{b}}} {Upward propagating high-frequency Alfv{\'e}n waves
  as identified from dynamic wave-like spicules observed by SOT on Hinode}.
\newblock \emph{Astron. Astrophys.}, \textbf{497}, 525--535.
\newblock (\doi{10.1051/0004-6361/200810777})

\bibitem[{{Heyvaerts} \& {Priest}(1983)}]{Heyvaerts1983}
{Heyvaerts}, J. \& {Priest}, E.~R. 1983 {Coronal heating by phase-mixed shear
  Alfven waves}.
\newblock \emph{Astron. Astrophys.}, \textbf{117}, 220--234.

\bibitem[{{Heyvaerts} \emph{et~al.}(1977){Heyvaerts}, {Priest} \&
  {Rust}}]{Heyvaerts1977}
{Heyvaerts}, J., {Priest}, E.~R. \& {Rust}, D.~M. 1977 {An emerging flux model
  for the solar flare phenomenon}.
\newblock \emph{Astrophys. J.}, \textbf{216}, 123--137.
\newblock (\doi{10.1086/155453})

\bibitem[{{Hollweg}(1978)}]{Hollweg1978}
{Hollweg}, J.~V. 1978 {Alfven waves in the solar atmosphere}.
\newblock \emph{Solar Phys.}, \textbf{56}, 305--333.
\newblock (\doi{10.1007/BF00152474})

\bibitem[{{Hollweg}(1984)}]{Hollweg1984}
{Hollweg}, J.~V. 1984 {Resonances of coronal loops}.
\newblock \emph{Astrophys. J.}, \textbf{277}, 392--403.
\newblock (\doi{10.1086/161706})

\bibitem[{{Hollweg} \& {Yang}(1988)}]{Hollweg1988}
{Hollweg}, J.~V. \& {Yang}, G. 1988 {Resonance absorption of compressible
  magnetohydrodynamic waves at thin 'surfaces'}.
\newblock \emph{J. Geophys. Res.}, \textbf{93}, 5423--5436.
\newblock (\doi{10.1029/JA093iA06p05423})

\bibitem[{{Hood} \emph{et~al.}(2009){Hood}, {Browning} \& {van der
  Linden}}]{Hoodetal2009}
{Hood}, A.~W., {Browning}, P.~K. \& {van der Linden}, R.~A.~M. 2009 {Coronal
  heating by magnetic reconnection in loops with zero net current}.
\newblock \emph{Astron. Astrophys.}, \textbf{506}, 913--925.
\newblock (\doi{10.1051/0004-6361/200912285})

\bibitem[{{Hudson}(1991)}]{Hudson1991}
{Hudson}, H.~S. 1991 {Solar flares, microflares, nanoflares, and coronal
  heating}.
\newblock \emph{Solar Phys.}, \textbf{133}, 357--369.
\newblock (\doi{10.1007/BF00149894})

\bibitem[{{Ionson}(1978)}]{Ionson1978}
{Ionson}, J.~A. 1978 {Resonant absorption of Alfvenic surface waves and the
  heating of solar coronal loops}.
\newblock \emph{Astrophys. J.}, \textbf{226}, 650--673.
\newblock (\doi{10.1086/156648})

\bibitem[{{Jefferies} \emph{et~al.}(2006){Jefferies}, {McIntosh}, {Armstrong},
  {Bogdan}, {Cacciani} \& {Fleck}}]{Jefferies2006}
{Jefferies}, S.~M., {McIntosh}, S.~W., {Armstrong}, J.~D., {Bogdan}, T.~J.,
  {Cacciani}, A. \& {Fleck}, B. 2006 {Magnetoacoustic Portals and the Basal
  Heating of the Solar Chromosphere}.
\newblock \emph{Astrophys. J. Lett.}, \textbf{648}, L151--L155.
\newblock (\doi{10.1086/508165})

\bibitem[{{Jess} \emph{et~al.}(2009){Jess}, {Mathioudakis}, {Erd{\'e}lyi},
  {Crockett}, {Keenan} \& {Christian}}]{Jess2009}
{Jess}, D.~B., {Mathioudakis}, M., {Erd{\'e}lyi}, R., {Crockett}, P.~J.,
  {Keenan}, F.~P. \& {Christian}, D.~J. 2009 {Alfv{\'e}n Waves in the Lower
  Solar Atmosphere}.
\newblock \emph{Science}, \textbf{323}, 1582--.
\newblock (\doi{10.1126/science.1168680})

\bibitem[{{Kano} \& {Tsuneta}(1995)}]{Kano1995}
{Kano}, R. \& {Tsuneta}, S. 1995 {Scaling Law of Solar Coronal Loops Obtained
  with YOHKOH}.
\newblock \emph{Astrophys. J.}, \textbf{454}, 934--+.
\newblock (\doi{10.1086/176547})

\bibitem[{{Klimchuk}(2006)}]{Klimchuk2006}
{Klimchuk}, J.~A. 2006 {On Solving the Coronal Heating Problem}.
\newblock \emph{Solar Phys.}, \textbf{234}, 41--77.
\newblock (\doi{10.1007/s11207-006-0055-z})

\bibitem[{{Klimchuk} \emph{et~al.}(2006){Klimchuk}, {L{\'o}pez Fuentes} \&
  {Devore}}]{Klimchuketal2006}
{Klimchuk}, J.~A., {L{\'o}pez Fuentes}, M.~C. \& {Devore}, C.~R. 2006 {Heating
  of the Magnetically Closed Corona}.
\newblock In \emph{Soho-17. 10 years of soho and beyond}, vol. 617 of \emph{ESA
  Special Publication}.

\bibitem[{{Krucker} \& {Benz}(1998)}]{Krucker1998}
{Krucker}, S. \& {Benz}, A.~O. 1998 {Energy Distribution of Heating Processes
  in the Quiet Solar Corona}.
\newblock \emph{Astrophys. J. Lett.}, \textbf{501}, L213+.
\newblock (\doi{10.1086/311474})

\bibitem[{{Kudoh} \& {Shibata}(1999)}]{Kudoh1999}
{Kudoh}, T. \& {Shibata}, K. 1999 {Alfv{\'e}n Wave Model of Spicules and
  Coronal Heating}.
\newblock \emph{Astrophys. J.}, \textbf{514}, 493--505.
\newblock (\doi{10.1086/306930})

\bibitem[{{Lenz} \emph{et~al.}(1999){Lenz}, {Deluca}, {Golub}, {Rosner} \&
  {Bookbinder}}]{Lenz1999}
{Lenz}, D.~D., {Deluca}, E.~E., {Golub}, L., {Rosner}, R. \& {Bookbinder},
  J.~A. 1999 {Temperature and Emission-Measure Profiles along Long-lived Solar
  Coronal Loops Observed with the Transition Region and Coronal Explorer}.
\newblock \emph{Astrophys. J. Lett.}, \textbf{517}, L155--L158.
\newblock (\doi{10.1086/312045})

\bibitem[{{Levine}(1974)}]{Levine1974}
{Levine}, R.~H. 1974 {A New Theory of Coronal Heating}.
\newblock \emph{Astrophys. J.}, \textbf{190}, 457--466.
\newblock (\doi{10.1086/152898})

\bibitem[{{Lin} \emph{et~al.}(1984){Lin}, {Schwartz}, {Kane}, {Pelling} \&
  {Hurley}}]{Lin1984}
{Lin}, R.~P., {Schwartz}, R.~A., {Kane}, S.~R., {Pelling}, R.~M. \& {Hurley},
  K.~C. 1984 {Solar hard X-ray microflares}.
\newblock \emph{Astrophys. J.}, \textbf{283}, 421--425.
\newblock (\doi{10.1086/162321})

\bibitem[{{Linsky}(1985)}]{linsky1985}
{Linsky}, J.~L. 1985 {Nonradiative activity across the H-R diagram - Which
  types of stars are solar-like?}
\newblock \emph{Solar Phys.}, \textbf{100}, 333--362.
\newblock (\doi{10.1007/BF00158435})

\bibitem[{{Longcope}(1998)}]{Longcope1998}
{Longcope}, D.~W. 1998 {A Model for Current Sheets and Reconnection in X-Ray
  Bright Points}.
\newblock \emph{Astrophys. J.}, \textbf{507}, 433--442.
\newblock (\doi{10.1086/306319})

\bibitem[{{Lundquist} \emph{et~al.}(2008){Lundquist}, {Fisher}, {Metcalf},
  {Leka} \& {McTiernan}}]{Lundquist2008}
{Lundquist}, L.~L., {Fisher}, G.~H., {Metcalf}, T.~R., {Leka}, K.~D. \&
  {McTiernan}, J.~M. 2008 {Forward Modeling of Active Region Coronal Emissions.
  II. Implications for Coronal Heating}.
\newblock \emph{Astrophys. J.}, \textbf{689}, 1388--1405.
\newblock (\doi{10.1086/592760})

\bibitem[{{Mart{\'{\i}}nez-Sykora} \emph{et~al.}(2009){Mart{\'{\i}}nez-Sykora},
  {Hansteen} \& {Carlsson}}]{Martinez-Sykora2009}
{Mart{\'{\i}}nez-Sykora}, J., {Hansteen}, V. \& {Carlsson}, M. 2009 {Twisted
  Flux Tube Emergence from the Convection Zone to the Corona. II. Later
  States}.
\newblock \emph{Astrophys. J.}, \textbf{702}, 129--140.
\newblock (\doi{10.1088/0004-637X/702/1/129})

\bibitem[{{Masuda} \emph{et~al.}(1994){Masuda}, {Kosugi}, {Hara}, {Tsuneta} \&
  {Ogawara}}]{Masuda1994}
{Masuda}, S., {Kosugi}, T., {Hara}, H., {Tsuneta}, S. \& {Ogawara}, Y. 1994 {A
  loop-top hard X-ray source in a compact solar flare as evidence for magnetic
  reconnection}.
\newblock \emph{Nature}, \textbf{371}, 495--497.
\newblock (\doi{10.1038/371495a0})

\bibitem[{{McIntosh} \emph{et~al.}(2011){McIntosh}, {de Pontieu}, {Carlsson},
  {Hansteen}, {Boerner} \& {Goossens}}]{McIntosh2011}
{McIntosh}, S.~W., {de Pontieu}, B., {Carlsson}, M., {Hansteen}, V., {Boerner},
  P. \& {Goossens}, M. 2011 {Alfv{\'e}nic waves with sufficient energy to power
  the quiet solar corona and fast solar wind}.
\newblock \emph{Nature}, \textbf{475}, 477--480.
\newblock (\doi{10.1038/nature10235})

\bibitem[{{McIntosh} \emph{et~al.}(2012){McIntosh}, {Tian}, {Sechler} \& {De
  Pontieu}}]{McIntosh2012}
{McIntosh}, S.~W., {Tian}, H., {Sechler}, M. \& {De Pontieu}, B. 2012 {On the
  Doppler Velocity of Emission Line Profiles Formed in the "Coronal Contraflow"
  that is the Chromospherer-Corona Mass Cycle}.
\newblock \emph{Astrophys. J. Lett.}, p. in press.

\bibitem[{{Mendoza-Brice{\~n}o} \& {Erd{\'e}lyi}(2006)}]{Mendoza2006}
{Mendoza-Brice{\~n}o}, C.~A. \& {Erd{\'e}lyi}, R. 2006 {Intermittent Coronal
  Loop Oscillations by Random Energy Releases}.
\newblock \emph{Astrophys. J.}, \textbf{648}, 722--731.
\newblock (\doi{10.1086/505642})

\bibitem[{{Mikic} \emph{et~al.}(1989){Mikic}, {Schnack} \& {van
  Hoven}}]{Mikic1989}
{Mikic}, Z., {Schnack}, D.~D. \& {van Hoven}, G. 1989 {Creation of current
  filaments in the solar corona}.
\newblock \emph{Astrophys. J.}, \textbf{338}, 1148--1157.
\newblock (\doi{10.1086/167265})

\bibitem[{{Moreno-Insertis} \emph{et~al.}(2008){Moreno-Insertis}, {Galsgaard}
  \& {Ugarte-Urra}}]{Moreno-Insertis2008}
{Moreno-Insertis}, F., {Galsgaard}, K. \& {Ugarte-Urra}, I. 2008 {Jets in
  Coronal Holes: Hinode Observations and Three-dimensional Computer Modeling}.
\newblock \emph{Astrophys. J. Lett.}, \textbf{673}, L211--L214.
\newblock (\doi{10.1086/527560})

\bibitem[{{Moriyasu} \emph{et~al.}(2004){Moriyasu}, {Kudoh}, {Yokoyama} \&
  {Shibata}}]{Moriyasu2004}
{Moriyasu}, S., {Kudoh}, T., {Yokoyama}, T. \& {Shibata}, K. 2004 {The
  Nonlinear Alfv{\'e}n Wave Model for Solar Coronal Heating and Nanoflares}.
\newblock \emph{Astrophys. J.}, \textbf{601}, L107--L110.
\newblock (\doi{10.1086/381779})

\bibitem[{{Narain} \& {Ulmschneider}(1996)}]{Narain1996}
{Narain}, U. \& {Ulmschneider}, P. 1996 {Chromospheric and Coronal Heating
  Mechanisms II}.
\newblock \emph{Space Sci. Rev.}, \textbf{75}, 453--509.
\newblock (\doi{10.1007/BF00833341})

\bibitem[{{Ofman} \emph{et~al.}(1998){Ofman}, {Klimchuk} \&
  {Davila}}]{Ofman1998}
{Ofman}, L., {Klimchuk}, J.~A. \& {Davila}, J.~M. 1998 {A Self-consistent Model
  for the Resonant Heating of Coronal Loops: The Effects of Coupling with the
  Chromosphere}.
\newblock \emph{Astrophys. J.}, \textbf{493}, 474--+.
\newblock (\doi{10.1086/305109})

\bibitem[{{Okamoto} \emph{et~al.}(2007){Okamoto}, {Tsuneta}, {Berger},
  {Ichimoto}, {Katsukawa}, {Lites}, {Nagata}, {Shibata}, {Shimizu}
  \emph{et~al.}}]{Okamoto2007}
{Okamoto}, T.~J., {Tsuneta}, S., {Berger}, T.~E., {Ichimoto}, K., {Katsukawa},
  Y., {Lites}, B.~W., {Nagata}, S., {Shibata}, K., {Shimizu}, T. \emph{et~al.}
  2007 {Coronal Transverse Magnetohydrodynamic Waves in a Solar Prominence}.
\newblock \emph{Science}, \textbf{318}, 1577--.
\newblock (\doi{10.1126/science.1145447})

\bibitem[{{Parker}(1972)}]{Parker1972}
{Parker}, E.~N. 1972 {Topological Dissipation and the Small-Scale Fields in
  Turbulent Gases}.
\newblock \emph{Astrophys. J.}, \textbf{174}, 499--+.
\newblock (\doi{10.1086/151512})

\bibitem[{{Parker}(1988)}]{Parker1988}
{Parker}, E.~N. 1988 {Nanoflares and the solar X-ray corona}.
\newblock \emph{Astrophys. J.}, \textbf{330}, 474--479.
\newblock (\doi{10.1086/166485})

\bibitem[{{Parnell}(2004)}]{Parnell2004}
{Parnell}, C.~E. 2004 {The Role of Dynamic Brightenings in Coronal Heating}.
\newblock In \emph{Soho 15 coronal heating} (ed. {R.~W.~Walsh, J.~Ireland,
  D.~Danesy, \& B.~Fleck}), vol. 575 of \emph{ESA Special Publication}, pp.
  227--+.

\bibitem[{{Parnell} \& {Haynes}(2010)}]{Parnell2010i}
{Parnell}, C.~E. \& {Haynes}, A.~L. 2010 {Three-Dimensional Magnetic
  Reconnection}.
\newblock In \emph{Magnetic coupling between the interior and atmosphere of the
  sun} (ed. {S.~S.~Hasan \& R.~J.~Rutten}), pp. 261--276.

\bibitem[{{Parnell} \emph{et~al.}(2008){Parnell}, {Haynes} \&
  {Galsgaard}}]{Parnell2008}
{Parnell}, C.~E., {Haynes}, A.~L. \& {Galsgaard}, K. 2008 {Recursive
  Reconnection and Magnetic Skeletons}.
\newblock \emph{Astrophys. J.}, \textbf{675}, 1656--1665.
\newblock (\doi{10.1086/527532})

\bibitem[{{Parnell} \& {Jupp}(2000)}]{Parnell2000}
{Parnell}, C.~E. \& {Jupp}, P.~E. 2000 {Statistical Analysis of the Energy
  Distribution of Nanoflares in the Quiet Sun}.
\newblock \emph{Astrophys. J.}, \textbf{529}, 554--569.
\newblock (\doi{10.1086/308271})

\bibitem[{{Parnell} \emph{et~al.}(2010){Parnell}, {Maclean} \&
  {Haynes}}]{Parnell2010}
{Parnell}, C.~E., {Maclean}, R.~C. \& {Haynes}, A.~L. 2010 {The Detection of
  Numerous Magnetic Separators in a Three-Dimensional Magnetohydrodynamic Model
  of Solar Emerging Flux}.
\newblock \emph{Astrophys. J. Lett.}, \textbf{725}, L214--L218.
\newblock (\doi{10.1088/2041-8205/725/2/L214})

\bibitem[{{Parnell} \emph{et~al.}(2011){Parnell}, {Maclean}, {Haynes} \&
  {Galsgaard}}]{Parnell2011}
{Parnell}, C.~E., {Maclean}, R.~C., {Haynes}, A.~L. \& {Galsgaard}, K. 2011 {3D
  Magnetic Reconnection}.
\newblock In \emph{Iau symposium}, vol. 271 of \emph{IAU Symposium}, pp.
  227--238.
\newblock (\doi{10.1017/S1743921311017650})

\bibitem[{{Parnell} \emph{et~al.}(1994){Parnell}, {Priest} \&
  {Golub}}]{Parnell1994}
{Parnell}, C.~E., {Priest}, E.~R. \& {Golub}, L. 1994 {The three-dimensional
  structures of X-ray bright points}.
\newblock \emph{Solar Phys.}, \textbf{151}, 57--74.
\newblock (\doi{10.1007/BF00654082})

\bibitem[{{Peter} \& {Judge}(1999)}]{PeterJudge1999}
{Peter}, H. \& {Judge}, P.~G. 1999 {On the Doppler Shifts of Solar Ultraviolet
  Emission Lines}.
\newblock \emph{Astrophys. J.}, \textbf{522}, 1148--1166.
\newblock (\doi{10.1086/307672})

\bibitem[{{Petschek}(1964)}]{Petschek1964}
{Petschek}, H.~E. 1964 {Magnetic Field Annihilation}.
\newblock \emph{NASA Special Publication}, \textbf{50}, 425.

\bibitem[{{Pevtsov} \emph{et~al.}(2003){Pevtsov}, {Fisher}, {Acton},
  {Longcope}, {Johns-Krull}, {Kankelborg} \& {Metcalf}}]{Pevtsov2003}
{Pevtsov}, A.~A., {Fisher}, G.~H., {Acton}, L.~W., {Longcope}, D.~W.,
  {Johns-Krull}, C.~M., {Kankelborg}, C.~C. \& {Metcalf}, T.~R. 2003 {The
  Relationship Between X-Ray Radiance and Magnetic Flux}.
\newblock \emph{Astrophys. J.}, \textbf{598}, 1387--1391.
\newblock (\doi{10.1086/378944})

\bibitem[{{Pontin} \emph{et~al.}(2011){Pontin}, {Wilmot-Smith}, {Hornig} \&
  {Galsgaard}}]{Pontin2011}
{Pontin}, D.~I., {Wilmot-Smith}, A.~L., {Hornig}, G. \& {Galsgaard}, K. 2011
  {Dynamics of braided coronal loops. II. Cascade to multiple small-scale
  reconnection events}.
\newblock \emph{Astron. Astrophys.}, \textbf{525}, A57+.
\newblock (\doi{10.1051/0004-6361/201014544})

\bibitem[{{Porter} \& {Klimchuk}(1995)}]{Porter1995}
{Porter}, L.~J. \& {Klimchuk}, J.~A. 1995 {Soft X-Ray Loops and Coronal
  Heating}.
\newblock \emph{Astrophys. J.}, \textbf{454}, 499--+.
\newblock (\doi{10.1086/176501})

\bibitem[{{Porter} \emph{et~al.}(1994){Porter}, {Klimchuk} \&
  {Sturrock}}]{Porter1994}
{Porter}, L.~J., {Klimchuk}, J.~A. \& {Sturrock}, P.~A. 1994 {The possible role
  of MHD waves in heating the solar corona}.
\newblock \emph{Astrophys. J.}, \textbf{435}, 482--501.
\newblock (\doi{10.1086/174830})

\bibitem[{{Priest} \& {Forbes}(2000)}]{Priest2000}
{Priest}, E. \& {Forbes}, T. 2000 \emph{{Magnetic Reconnection}}.
\newblock Cambridge University Press, Cambridge, UK.

\bibitem[{{Priest}(1982)}]{Priest1982}
{Priest}, E.~R. 1982 \emph{{Solar magneto-hydrodynamics}}.
\newblock Dordrecht, Holland ; Boston : D.~Reidel Pub.~Co.~; Hingham,.

\bibitem[{{Priest} \emph{et~al.}(2002){Priest}, {Heyvaerts} \&
  {Title}}]{Priest2002}
{Priest}, E.~R., {Heyvaerts}, J.~F. \& {Title}, A.~M. 2002 {A Flux-Tube
  Tectonics Model for Solar Coronal Heating Driven by the Magnetic Carpet}.
\newblock \emph{Astrophys. J.}, \textbf{576}, 533--551.
\newblock (\doi{10.1086/341539})

\bibitem[{{Priest} \emph{et~al.}(1994){Priest}, {Parnell} \&
  {Martin}}]{Priest1994}
{Priest}, E.~R., {Parnell}, C.~E. \& {Martin}, S.~F. 1994 {A converging flux
  model of an X-ray bright point and an associated canceling magnetic feature}.
\newblock \emph{Astrophys. J.}, \textbf{427}, 459--474.
\newblock (\doi{10.1086/174157})

\bibitem[{{Rae} \& {Roberts}(1981)}]{Rae1981}
{Rae}, I.~C. \& {Roberts}, B. 1981 {Surface waves and the heating of the
  corona}.
\newblock \emph{Geophysical and Astrophysical Fluid Dynamics}, \textbf{18},
  197--226.
\newblock (\doi{10.1080/03091928108208836})

\bibitem[{{Rappazzo} \emph{et~al.}(2010){Rappazzo}, {Velli} \&
  {Einaudi}}]{Rappazzoetal2010}
{Rappazzo}, A.~F., {Velli}, M. \& {Einaudi}, G. 2010 {Shear Photospheric
  Forcing and the Origin of Turbulence in Coronal Loops}.
\newblock \emph{Astrophys. J.}, \textbf{722}, 65--78.
\newblock (\doi{10.1088/0004-637X/722/1/65})

\bibitem[{{Rappazzo} \emph{et~al.}(2007){Rappazzo}, {Velli}, {Einaudi} \&
  {Dahlburg}}]{Rappazzoetal2007}
{Rappazzo}, A.~F., {Velli}, M., {Einaudi}, G. \& {Dahlburg}, R.~B. 2007
  {Coronal Heating, Weak MHD Turbulence, and Scaling Laws}.
\newblock \emph{Astrophys. J. Lett.}, \textbf{657}, L47--L51.
\newblock (\doi{10.1086/512975})

\bibitem[{{Reale}(2007)}]{Reale2007}
{Reale}, F. 2007 {Diagnostics of stellar flares from X-ray observations: from
  the decay to the rise phase}.
\newblock \emph{Astron. Astrophys.}, \textbf{471}, 271--279.
\newblock (\doi{10.1051/0004-6361:20077223})

\bibitem[{{Reale}(2010)}]{Reale2010}
{Reale}, F. 2010 {Coronal Loops: Observations and Modeling of Confined Plasma}.
\newblock \emph{Living Reviews in Solar Physics}, \textbf{7}, 5--+.

\bibitem[{{Reale} \emph{et~al.}(2011){Reale}, {Guarrasi}, {Testa}, {DeLuca},
  {Peres} \& {Golub}}]{Reale2011}
{Reale}, F., {Guarrasi}, M., {Testa}, P., {DeLuca}, E.~E., {Peres}, G. \&
  {Golub}, L. 2011 {Solar Dynamics Observatory Discovers Thin High Temperature
  Strands in Coronal Active Regions}.
\newblock \emph{Astrophys. J. Lett.}, \textbf{736}, L16+.
\newblock (\doi{10.1088/2041-8205/736/1/L16})

\bibitem[{{Rosenthal} \emph{et~al.}(2002){Rosenthal}, {Bogdan}, {Carlsson},
  {Dorch}, {Hansteen}, {McIntosh}, {McMurry}, {Nordlund} \&
  {Stein}}]{Rosenthal2002}
{Rosenthal}, C.~S., {Bogdan}, T.~J., {Carlsson}, M., {Dorch}, S.~B.~F.,
  {Hansteen}, V., {McIntosh}, S.~W., {McMurry}, A., {Nordlund}, {\AA}. \&
  {Stein}, R.~F. 2002 {Waves in the Magnetized Solar Atmosphere. I. Basic
  Processes and Internetwork Oscillations}.
\newblock \emph{Astrophys. J.}, \textbf{564}, 508--524.
\newblock (\doi{10.1086/324214})

\bibitem[{{Rosner} \emph{et~al.}(1978){Rosner}, {Tucker} \&
  {Vaiana}}]{Rosner1978}
{Rosner}, R., {Tucker}, W.~H. \& {Vaiana}, G.~S. 1978 {Dynamics of the
  quiescent solar corona}.
\newblock \emph{Astrophys. J.}, \textbf{220}, 643--645.
\newblock (\doi{10.1086/155949})

\bibitem[{{Ruderman} \emph{et~al.}(1997{\natexlab{\emph{a}}}){Ruderman},
  {Berghmans}, {Goossens} \& {Poedts}}]{Ruderman1997a}
{Ruderman}, M.~S., {Berghmans}, D., {Goossens}, M. \& {Poedts}, S.
  1997{\natexlab{\emph{a}}} {Direct excitation of resonant torsional Alfven
  waves by footpoint motions.}
\newblock \emph{Astron. Astrophys.}, \textbf{320}, 305--318.

\bibitem[{{Ruderman} \emph{et~al.}(1997{\natexlab{\emph{b}}}){Ruderman},
  {Goossens}, {Ballester} \& {Oliver}}]{Ruderman1997b}
{Ruderman}, M.~S., {Goossens}, M., {Ballester}, J.~L. \& {Oliver}, R.
  1997{\natexlab{\emph{b}}} {Resonant Alfven waves in coronal arcades driven by
  footpoint motions}.
\newblock \emph{Astron. Astrophys.}, \textbf{328}, 361--370.

\bibitem[{{Sakamoto} \emph{et~al.}(2008){Sakamoto}, {Tsuneta} \&
  {Vekstein}}]{Sakamoto2008}
{Sakamoto}, Y., {Tsuneta}, S. \& {Vekstein}, G. 2008 {Observational Appearance
  of Nanoflares with SXT and TRACE}.
\newblock \emph{Astrophys. J.}, \textbf{689}, 1421--1432.
\newblock (\doi{10.1086/592488})

\bibitem[{{Schmelz} \emph{et~al.}(2009){Schmelz}, {Saar}, {DeLuca}, {Golub},
  {Kashyap}, {Weber} \& {Klimchuk}}]{Schmeltz2009}
{Schmelz}, J.~T., {Saar}, S.~H., {DeLuca}, E.~E., {Golub}, L., {Kashyap},
  V.~L., {Weber}, M.~A. \& {Klimchuk}, J.~A. 2009 {Hinode X-Ray Telescope
  Detection of Hot Emission from Quiescent Active Regions: A Nanoflare
  Signature?}
\newblock \emph{Astrophys. J. Lett.}, \textbf{693}, L131--L135.
\newblock (\doi{10.1088/0004-637X/693/2/L131})

\bibitem[{{Schrijver} \emph{et~al.}(2004){Schrijver}, {Sandman}, {Aschwanden}
  \& {De Rosa}}]{Schrijver2004}
{Schrijver}, C.~J., {Sandman}, A.~W., {Aschwanden}, M.~J. \& {De Rosa}, M.~L.
  2004 {The Coronal Heating Mechanism as Identified by Full-Sun
  Visualizations}.
\newblock \emph{Astrophys. J.}, \textbf{615}, 512--525.
\newblock (\doi{10.1086/424028})

\bibitem[{{Schwarzschild}(1948)}]{Schwarzschild1948}
{Schwarzschild}, M. 1948 {On Noise Arising from the Solar Granulation.}
\newblock \emph{Astrophys. J.}, \textbf{107}, 1--+.
\newblock (\doi{10.1086/144983})

\bibitem[{{Shibata} \emph{et~al.}(1992{\natexlab{\emph{a}}}){Shibata},
  {Ishido}, {Acton}, {Strong}, {Hirayama}, {Uchida}, {McAllister}, {Matsumoto},
  {Tsuneta} \emph{et~al.}}]{Shibata1992a}
{Shibata}, K., {Ishido}, Y., {Acton}, L.~W., {Strong}, K.~T., {Hirayama}, T.,
  {Uchida}, Y., {McAllister}, A.~H., {Matsumoto}, R., {Tsuneta}, S.
  \emph{et~al.} 1992{\natexlab{\emph{a}}} {Observations of X-ray jets with the
  YOHKOH Soft X-ray Telescope}.
\newblock \emph{Pub.\ Astron.\ Soc.\ J.}, \textbf{44}, L173--L179.

\bibitem[{{Shibata} \emph{et~al.}(1992{\natexlab{\emph{b}}}){Shibata}, {Nozawa}
  \& {Matsumoto}}]{Shibata1992b}
{Shibata}, K., {Nozawa}, S. \& {Matsumoto}, R. 1992{\natexlab{\emph{b}}}
  {Magnetic reconnection associated with emerging magnetic flux}.
\newblock \emph{Pub.\ Astron.\ Soc.\ J.}, \textbf{44}, 265--272.

\bibitem[{{Shimizu}(1995)}]{Shimizu1995}
{Shimizu}, T. 1995 {Energetics and Occurrence Rate of Active-Region Transient
  Brightenings and Implications for the Heating of the Active-Region Corona}.
\newblock \emph{Pub. Astron. Soc. Japan}, \textbf{47}, 251--263.

\bibitem[{{Taroyan} \& {Erd{\'e}lyi}(2009)}]{Taroyan2009}
{Taroyan}, Y. \& {Erd{\'e}lyi}, R. 2009 {Heating Diagnostics with MHD Waves}.
\newblock \emph{Space Science Rev.}, \textbf{149}, 229--254.
\newblock (\doi{10.1007/s11214-009-9506-9})

\bibitem[{{Taroyan} \emph{et~al.}(2007){Taroyan}, {Erd{\'e}lyi}, {Doyle} \&
  {Bradshaw}}]{Taroyan2007}
{Taroyan}, Y., {Erd{\'e}lyi}, R., {Doyle}, J.~G. \& {Bradshaw}, S.~J. 2007
  {Analysis of power spectra of Doppler shift time series as a diagnostic tool
  for quiescent coronal loops}.
\newblock \emph{Astron. Astrophys.}, \textbf{462}, 331--340.
\newblock (\doi{10.1051/0004-6361:20066069})

\bibitem[{{Terzo} \emph{et~al.}(2011){Terzo}, {Reale}, {Miceli}, {Klimchuk},
  {Kano} \& {Tsuneta}}]{Terzo2011}
{Terzo}, S., {Reale}, F., {Miceli}, M., {Klimchuk}, J.~A., {Kano}, R. \&
  {Tsuneta}, S. 2011 {Widespread Nanoflare Variability Detected with
  Hinode/X-Ray Telescope in a Solar Active Region}.
\newblock \emph{Astrophys. J.}, \textbf{736}, 111.
\newblock (\doi{10.1088/0004-637X/736/2/111})

\bibitem[{{Thornton} \& {Parnell}(2011)}]{Thornton2011}
{Thornton}, L.~M. \& {Parnell}, C.~E. 2011 {Small-Scale Flux Emergence Observed
  Using Hinode/SOT}.
\newblock \emph{Solar Phys.}, \textbf{269}, 13--40.
\newblock (\doi{10.1007/s11207-010-9656-7})

\bibitem[{{Tomczyk} \& {McIntosh}(2009)}]{Tomczyk2009}
{Tomczyk}, S. \& {McIntosh}, S.~W. 2009 {Time-Distance Seismology of the Solar
  Corona with CoMP}.
\newblock \emph{Astrophys. J.}, \textbf{697}, 1384--1391.
\newblock (\doi{10.1088/0004-637X/697/2/1384})

\bibitem[{{Tomczyk} \emph{et~al.}(2007){Tomczyk}, {McIntosh}, {Keil}, {Judge},
  {Schad}, {Seeley} \& {Edmondson}}]{Tomczyk2007}
{Tomczyk}, S., {McIntosh}, S.~W., {Keil}, S.~L., {Judge}, P.~G., {Schad}, T.,
  {Seeley}, D.~H. \& {Edmondson}, J. 2007 {Alfv{\'e}n Waves in the Solar
  Corona}.
\newblock \emph{Science}, \textbf{317}, 1192--.
\newblock (\doi{10.1126/science.1143304})

\bibitem[{{Vaiana}(1981)}]{vaiana1981}
{Vaiana}, G.~S. 1981 {The Einstein/CFA stellar survey - Overview of the data
  and interpretation of results}.
\newblock In \emph{X-ray astronomy with the einstein satellite} (ed.
  {R.~Giacconi}), vol.~87 of \emph{Astrophysics and Space Science Library}, pp.
  1--18.

\bibitem[{{Van Doorsselaere} \emph{et~al.}(2007){Van Doorsselaere}, {Andries}
  \& {Poedts}}]{VanDoorsselaere2007}
{Van Doorsselaere}, T., {Andries}, J. \& {Poedts}, S. 2007 {Observational
  evidence favors a resistive wave heating mechanism for coronal loops over a
  viscous phenomenon}.
\newblock \emph{Astron. Astrophys.}, \textbf{471}, 311--314.
\newblock (\doi{10.1051/0004-6361:20066658})

\bibitem[{{Verwichte} \emph{et~al.}(2010){Verwichte}, {Marsh}, {Foullon}, {Van
  Doorsselaere}, {De Moortel}, {Hood} \& {Nakariakov}}]{Verwichte2010}
{Verwichte}, E., {Marsh}, M., {Foullon}, C., {Van Doorsselaere}, T., {De
  Moortel}, I., {Hood}, A.~W. \& {Nakariakov}, V.~M. 2010 {Periodic Spectral
  Line Asymmetries in Solar Coronal Structures from Slow Magnetoacoustic
  Waves}.
\newblock \emph{Astrophys. J. Lett.}, \textbf{724}, L194--L198.
\newblock (\doi{10.1088/2041-8205/724/2/L194})

\bibitem[{{von Rekowski} \& {Hood}(2008)}]{vonRekowski2008}
{von Rekowski}, B. \& {Hood}, A.~W. 2008 {Photospheric cancelling magnetic
  features and associated phenomena in a stratified solar atmosphere}.
\newblock \emph{MNRAS}, \textbf{385}, 1792--1812.
\newblock (\doi{10.1111/j.1365-2966.2008.12968.x})

\bibitem[{{Warren} \emph{et~al.}(2008){Warren}, {Ugarte-Urra}, {Doschek},
  {Brooks} \& {Williams}}]{Warren2008a}
{Warren}, H.~P., {Ugarte-Urra}, I., {Doschek}, G.~A., {Brooks}, D.~H. \&
  {Williams}, D.~R. 2008 {Observations of Active Region Loops with the EUV
  Imaging Spectrometer on Hinode}.
\newblock \emph{Asrophys. J. Lett.}, \textbf{686}, L131--L134.
\newblock (\doi{10.1086/592960})

\bibitem[{{Warren} \& {Winebarger}(2006)}]{Warren2006}
{Warren}, H.~P. \& {Winebarger}, A.~R. 2006 {Hydrostatic Modeling of the
  Integrated Soft X-Ray and Extreme Ultraviolet Emission in Solar Active
  Regions}.
\newblock \emph{Astrophys. J.}, \textbf{645}, 711--719.
\newblock (\doi{10.1086/504075})

\bibitem[{{Warren} \emph{et~al.}(2010){Warren}, {Winebarger} \&
  {Brooks}}]{Warren2010}
{Warren}, H.~P., {Winebarger}, A.~R. \& {Brooks}, D.~H. 2010 {Evidence for
  Steady Heating: Observations of an Active Region Core with Hinode and TRACE}.
\newblock \emph{Astrophys. J.}, \textbf{711}, 228--238.
\newblock (\doi{10.1088/0004-637X/711/1/228})

\bibitem[{{Warren} \emph{et~al.}(2002){Warren}, {Winebarger} \&
  {Hamilton}}]{Warren2002}
{Warren}, H.~P., {Winebarger}, A.~R. \& {Hamilton}, P.~S. 2002 {Hydrodynamic
  Modeling of Active Region Loops}.
\newblock \emph{Asrophys. J. Lett.}, \textbf{579}, L41--L44.
\newblock (\doi{10.1086/344921})

\bibitem[{{Warren} \emph{et~al.}(2003){Warren}, {Winebarger} \&
  {Mariska}}]{Warren2003}
{Warren}, H.~P., {Winebarger}, A.~R. \& {Mariska}, J.~T. 2003 {Evolving Active
  Region Loops Observed with the Transition Region and Coronal explorer. II.
  Time-dependent Hydrodynamic Simulations}.
\newblock \emph{Astrophys. J.}, \textbf{593}, 1174--1186.
\newblock (\doi{10.1086/376678})

\bibitem[{{Wentzel}(1974)}]{Wentzel1974}
{Wentzel}, D.~G. 1974 {Coronal heating by Alfven waves}.
\newblock \emph{Solar Phys.}, \textbf{39}, 129--140.
\newblock (\doi{10.1007/BF00154975})

\bibitem[{{Wentzel}(1976)}]{Wentzel1976}
{Wentzel}, D.~G. 1976 {Coronal heating by Alfven waves. II}.
\newblock \emph{Solar Phys.}, \textbf{50}, 343--360.
\newblock (\doi{10.1007/BF00155297})

\bibitem[{{Wilmot-Smith} \emph{et~al.}(2010){Wilmot-Smith}, {Pontin} \&
  {Hornig}}]{Wilmot-Smith2010}
{Wilmot-Smith}, A.~L., {Pontin}, D.~I. \& {Hornig}, G. 2010 {Dynamics of
  braided coronal loops. I. Onset of magnetic reconnection}.
\newblock \emph{Astron. Astrophys.}, \textbf{516}, A5+.
\newblock (\doi{10.1051/0004-6361/201014041})

\bibitem[{{Winebarger} \emph{et~al.}(2008){Winebarger}, {Warren} \&
  {Falconer}}]{Winebarger2008}
{Winebarger}, A.~R., {Warren}, H.~P. \& {Falconer}, D.~A. 2008 {Modeling X-Ray
  Loops and EUV ``Moss'' in an Active Region Core}.
\newblock \emph{Astrophys. J.}, \textbf{676}, 672--679.
\newblock (\doi{10.1086/527291})

\bibitem[{{Winebarger} \emph{et~al.}(2003){Winebarger}, {Warren} \&
  {Mariska}}]{Winebarger2003}
{Winebarger}, A.~R., {Warren}, H.~P. \& {Mariska}, J.~T. 2003 {Transition
  Region and Coronal Explorer and Soft X-Ray Telescope Active Region Loop
  Observations: Comparisons with Static Solutions of the Hydrodynamic
  Equations}.
\newblock \emph{Astrophys. J.}, \textbf{587}, 439--449.
\newblock (\doi{10.1086/368017})

\bibitem[{{Withbroe} \& {Noyes}(1977)}]{Withbroe1977}
{Withbroe}, G.~L. \& {Noyes}, R.~W. 1977 {Mass and energy flow in the solar
  chromosphere and corona}.
\newblock \emph{Ann. Rev. Astron. Astrophys.}, \textbf{15}, 363--387.
\newblock (\doi{10.1146/annurev.aa.15.090177.002051})

\bibitem[{{Yokoyama} \& {Shibata}(1995)}]{Yokoyama1995}
{Yokoyama}, T. \& {Shibata}, K. 1995 {Magnetic reconnection as the origin of
  X-ray jets and H{$\alpha$} surges on the Sun}.
\newblock \emph{Nature}, \textbf{375}, 42--44.
\newblock (\doi{10.1038/375042a0})

\end{thebibliography}

\end{document}